\documentclass[twocolumn,preprintnumbers,amsmath,amssymb,showpacs]{revtex4}

\usepackage{graphicx}
\usepackage{dcolumn}
\usepackage{bm}

\newcommand{\beq}{\begin{equation}}
\newcommand{\eeq}{\end{equation}}
\newcommand{\barr}{\begin{eqnarray}}
\newcommand{\earr}{\end{eqnarray}}
\newcommand{\bra}[1]{\left\langle #1 \right|}
\newcommand{\ket}[1]{\left| #1 \right\rangle}
\newcommand{\de}[2]{\frac{\partial #1}{\partial #2}}

\newcommand{\é}{\'e}

\newcommand{\à}{\`a}

\newcommand{\dominioRtau}{\mathbb{R}^n \times[0,\tau]}
\newcommand{\dpr}[1]{\partial_{ #1 }}
\newcommand{\dtot}[2]{\frac{d #1}{d #2}}
\newcommand{\rif}[1]{(\ref{eq#1})}

\begin{document}

\title{Quantum Theory of particles and fields as an extension of a probabilistic variational approach to classical mechanics and classical field theory. I}
\author{M. Villani}
\affiliation{Dipartimento di Fisica - Universit\à di Bari - Italy
\\ Istituto Nazionale di Fisica Nucleare - Sezione di Bari}
\email{matteo.villani@ba.infn.it}
\date{July 2009 (revised version)}

\begin{abstract}
A theoretical scheme, based on a probabilistic generalization of the Hamilton's principle, is
elaborated to obtain an unified description of more general dynamical behaviors determined both
from a lagrangian function and by mechanisms not contemplated by this function. Within this
scheme, quantum mechanics, classical field theory and a quantum theory for scalar fields are
discussed. As a by-product of the probabilistic scheme for classical field theory, the
equations of the De Donder-Weyl theory for multi-dimensional variational problems are
recovered.
\end{abstract}

\maketitle

\section{\label{sec:level1}Introduction}

The main purpose of this paper is to explore the possibility of reconsidering the problem of
the relativistic quantum theory of fields from a different point of view. We will focus on
standards fields with no reference to strings.

Let us start from the following general remarks. Suppose to have a classical isolated physical
system described by a set of real space-time functions (fields) $ q^i (x) (i = 1,2,..,n), x =
(x^0,x^1,x^2,x^3) \in M$, the space-time manifold, with the metric tensor
$g_{\mu\nu}=diag(1,-1,-1,-1)$. If it is associated to the system a Lagrangian (density)

\beq \label{eq1} \mathcal{L} = \mathcal{L}(q^i(x),\partial_\mu
q^i(x)) \qquad (\partial_\mu = \frac{\partial}{\partial x^\mu} ,
\mu = 0,1,2,3)  \eeq

then, as is well known, the field equations determining the
functions $q^i(x)$ can be obtained through a variational
principle, that is by requiring that the action

\beq  \label{eq2} A_c = \int_V \mathcal{L}(q^i(x),\partial_\mu
q^i(x)) dx \qquad  (dx = dx^0dx^1dx^2dx^3) \eeq

be stationary under variations of $q^i(x)$ vanishing at the boundary of V, a domain of M. Then
one obtains the partial differential (Euler-Lagrange) equations

\beq \label{eq3}  \partial_\mu (
\frac{\mathcal{L}}{\partial(\partial_\mu q^i(x))}) -
\frac{\partial\mathcal{L}}{\partial q^i(x)} = 0 \qquad
(i=1,2,..,n) \eeq

In the case  of a classical mechanical system with a finite number of degrees of freedom
(m.s.f.), a useful and relevant tool for the study of its dynamical properties is given by the
Hamiltonian approach. There is essentially one Hamiltonian formulation associated to a given
Lagrangian. This formulation has a fundamental role in the transition from classical to quantum
theory, if this transition is performed on the basis of the standard canonical quantization
rules (s.c.q.r). But it is also very relevant and effective in conjunction with the
Hamilton-Jacobi theory and the general aspects of the one-dimensional variational problem
involved in the case of a m.s.f. \cite{bib_1,bib_2,bib_3}.

On the other hand, in the case of continuous systems described by fields, we do not have a
univocal Hamiltonian formulation associated to a given classical Lagrangian. The
multi-dimensional variational problem (\ref{eq2}) is tackled essentially according two
different points of view: one developed mainly in physics, the other in mathematics.

In physics, a continuous system is interpreted as a mechanical system with an infinite number
of degrees of freedom (m.s.i) \cite{bib_4,bib_5}. The spatial coordinates $(x^1,x^2,x^3)$ have
the role of labels of these infinite degrees of freedom. The Hamiltonian formalism is developed
on the basis of this mechanical transcription. Such a formulation has a crucial role in the
transition to the quantum theory of a continuous system, since it allows to apply the s.c.q.r.
of quantum mechanics. However it has a draw-back in the fundamental distinction between time
and space variables. As is well known, this problem is bypassed in quantum theory, by working
in the Heisemberg or Interaction picture, or in the framework of the Feynman's functional
approach. But, due to the original point of view, further problems appear, like the occurrence
of infinite quantities. In some cases these problems are bypassed through additional
prescriptions, like the renormalization. In these cases, the resulting (effective) quantum
field theory has been very successful. It has reached the stage of a paradigm, although
accompanied by doubts about its completeness. It runs into difficulties in the case of the
gravitational fields.

At a classical level, a draw-back of the mechanical transcription of a continuous system is
that it leads to an Hamilton-Jacobi equation formally of functional type, which frustrates its
use. On the other hand, the mathematical analysis of multi-dimensional variational problems
does not make an essential use of the interpretation of a classical continuous system as a
m.s.i. \cite{bib_1,bib_2,bib_6}. In this analysis, the space-coordinates are treated on the
same footing of the time: we have four parameters which together take the place that the time
alone has in a classical m.s.f. As a matter of fact we have several approaches to
multidimensional variational problems \cite{bib_1,bib_2,bib_6}, which share this point of view.
However one relevant approach has been formulated, called the De Donder-Weyl Theory
\cite{bib_1,bib_2,bib_6,bib_7}, which introduces and makes use of an "Hamiltion" function. In
this theory a covariant Hamiltonian formalism is elaborated \cite{bib_8} and a generalized
Hamilton-Jacobi equation is obtained, which is not a functional, but a partial differential
equation.

At the classical level, the De Donder-Weyl formalism appears more appropriate than the
Hamiltonian formalism based on the mechanical transcription. However, if we consider the
transition to a quantum theory, it is in conflict with the s.c.q.r. . In fact these rules
require only one canonical momentum (density) associated to each field $q^i(x)$, that is
$\partial\mathcal{L}/\partial_0q^i(x)$ (as happens in the theory based on the mechanical
analogy), while in the De Donder-Weyl theory we have four momenta \beq \label{eq4} \pi_i^\mu(x)
= \frac{\partial\mathcal{L}}{\partial_\mu q^i(x)} \qquad
 (\mu = 0,1,2,3; i=1,2,..,n) \eeq Several attempts have been
made in order to overcome the above difficulty and to obtain a covariant canonical quantization
\cite[and reference therein]{bib_9,bib_10,bib_11}. Unfortunately, in these attempts, there is
some arbitrariness and some insubstantiality in the probabilistic interpretation.

It seems that the mechanical point of view is necessary, if one insists on the s.c.q.r. as the
unique tool for a quantum theory of fields. In the past there are been objections to this point
of view \cite{bib_12,bib_13}. As a matter of fact, Born \cite{bib_13} tried to avoid the
s.c.q.r. and to introduce an alternative quantization procedure based on a result of the
multi-dimensional calculus of variations (the Hilbert independence theorem) \cite{bib_14}. But
his attempt was not successful.

In this paper we make again an attempt like that of Born, but firstly we elaborate an
appropriate alternative theoretical scheme for the analysis of the transition from classical
theory to quantum theory. Developing an approach proposed in a previous work \cite{bib_15}, we
adopt at the start, that is already at the classical level, a probabilistic point of view. In
this setting we make a systematic use of the variational approach, by introducing a
generalization of the Hamilton's principle. We obtain in this way a scheme more flexible and
more directly open to possible extensions.

The preliminary aspects of this scheme are discussed in Sect.\ref{sec:level2}, in the context
of classical mechanics. The transition to quantum mechanics is analyzed in
Sect.\ref{sec:level3}; this transition appears more as a natural extension than a conceptual
jump. A particular role is covered by time-reversal invariance and by the local or global
character of the dynamical equations. We will deduce the request for a mechanical system to be
natural and discuss a mechanism for the emergence of complex amplitudes. The case of a discrete
random variable (spin) is also analyzed. The probabilistic point of view is applied to
classical fields in Sect.\ref{sec:level4}. We avoid the mechanical transcription, since it
leads to an ill-defined probabilistic scheme. However we show that there exists a well defined
probabilistic scheme, in which the usual deterministic classical field theory can be embedded.
It is very interesting that, in this scheme, there appear, as a by-product, the equations of
the De Donder-Weyl theory. We see that the two Hamiltonian formulations, which are "equivalent"
from the deterministic point of view (in the sense that they give the same equations
(\ref{eq3})), have different implications if we adopt a probabilistic point of view.

The transition to a quantum theory of fields is discussed in Sect.\ref{sec:level5}. This
transition, in which the s.c.q.r. are avoided, is an extension of the results of the sections
\ref{sec:level3} and \ref{sec:level4}.

\section{\label{sec:level2}Probabilistic variational approach to classical mechanics}

We consider a conservative mechanical system with $n$ degrees of freedom and configuration
space $\mathbb{R}^n$, described by a Lagrangian $L$, i. e. a given smooth function of $2n$
independent variables \beq \label{eq5} L = L(q,w) \qquad (q=(q^1,q^2,..,q^n) ,
w=(w^1,w^2,..,w^n)) \eeq
 The coordinates $q$ give the configurations of the system. Let us fix
a time $\tau > 0$ and make in (\ref{eq5}) the substitutions \beq \label{eq6} q \rightarrow
\xi(t) \, , \, w \rightarrow \dot{\xi}(t) \qquad
 (\:\dot{}\,=\frac{d}{dt}) \eeq Then we can consider the
action functional \beq \label{eq7} I = \int_0^\tau L(\xi(t),\dot{\xi}(t))dt \eeq According to
the Hamilton's principle, the temporal evolution of our system is describes by its trajectories
$q(t)$, which are functions of $t$ such that \beq \label{eq8}
{\frac{d}{d\varepsilon}{I(\varepsilon)}}\biggr\lvert_{\epsilon=0} =
\frac{d}{d\varepsilon}\int_0^\tau
L(q(t)+\varepsilon\eta(t),\dot{q}(t)+\varepsilon\dot{\eta}(t))dt \biggr\rvert_{\epsilon=0} = 0
\eeq for every smooth function $\eta(t)$ having support in $[0,\tau]$. As is well known we
obtain in this way the Lagrange equations for $q(t)$ \beq \label{eq9} \frac{d}{dt}
\frac{\partial L}{\partial w^i} (q(t),\dot{q}(t))-\frac{\partial L}{\partial
q^i}(q(t),\dot{q}(t)) = 0 \qquad  (i=1,2,..\,,n) \eeq Now, we adopt at the start a
probabilistic point of view. At each instant $t \in [0,\tau]$, we consider the configuration
$q$ a random variable described by a probability density $\rho(q,t)$. We limit ourselves only
to this assumption, that is we do not require a priori that the temporal evolution is
necessarily described by a stochastic process. We will take into account of the obvious fact
that the total probability \beq \int_{\mathbb{R}^n} \rho(q,t)dq \qquad  (dq = dq^1..\,dq^n)
\nonumber \eeq does not depend of the time $t$, assuming that this conservation law can be
expressed in a local form \beq \label{eq10} \partial_t \rho(q,t)+\partial_k j^k(q,t) = 0\, ,
\qquad  (\partial_k = \frac{\partial}{\partial q^k})\eeq
through a probability current density $j(q,t) = ( j^1(q,t),\\
..\,,j^n(q,t))$.

We will consider $j(q,t)$ as a further dynamical variable, which, together with $\rho(q,t)$,
characterizes the "state" of the system. As a matter of fact, (\ref{eq10}) takes the place or
generalizes the link between $q(t)$ and $\dot{q}(t)$ of the standard theory based on the notion
of trajectory. We plan to determine (adding initial conditions) $\rho(q,t)$ and $j(q,t)$
directly from the given Lagrangian (\ref{eq5}), through an appropriate generalization of the
Hamilton's principle \cite{bib_15}. To this end, suppose that $\rho(q,t)$ and $j(q,t)$ describe
an actual temporal evolution of our system and let us consider the family of subsets $\Sigma =
\mathrm{U}\times[t_0,t_1]$, where $\mathrm{U}$ is a compact domain of $\mathbb{R}^n$ and $0\leq
t_0<t_1\leq \tau$, such that $\rho(q,t)>0$ for $q,t \in \Sigma$. In each $\Sigma$ let us make
the substitution
 \beq \label{eq11} w \rightarrow \: \frac{j(q,t)}{\rho(q,t)} \eeq
  in the given Lagrangian (\ref{eq5}). Due to the meaning of $j(q,t)$ and $\rho(q,t)$,
 this substitution is a generalization of (\ref{eq6}). Then let us consider the action functional
  \barr \label{eq12} I_\Sigma[j,\rho,\lambda]&=&\int_\Sigma
dtdq\biggr[L\bigr(q,\frac{j(q,t)}{\rho(q,t)}\bigr)\rho(q,t)+\nonumber
\\ &+& \lambda(q,t)\bigr(\partial_t \rho(q,t)+\partial_k
j^k(q,t)\bigr)\biggr],\earr where $\lambda(q,t)$ is a Lagrange multiplier field defined in
$\Sigma$.

We can now formulate a generalized Hamilton's principle, which replaces
(\ref{eq8}), as follows:\\
\begin{em}
"The actual functions $\rho(q,t)$ and $j(q,t)$ are such that the equation (\ref{eq10}) holds in
$\mathbb{R}^n \times[0,\tau]$, and in each $\Sigma$ there exists a function $\lambda(q,t)$ such
that \beq \label{eq13} \frac{\partial}{\partial \varepsilon_i}I_\Sigma \Bigr[j+\varepsilon_1
j_0,\rho + \varepsilon_2
\rho_0,\lambda\Bigr]\Biggr|_{\stackrel{\varepsilon_1=\,0}{\varepsilon_2=\,0}} = 0 \quad
(i=1,2) \eeq for every smooth function $j_0(q,t)$ for which $j_0(q,t)=0$ for $q\in \partial
\mathrm{U}$, and every smooth function $\rho_0(q,t)$ such that $\rho_0(q,t_0)=\rho_0(q,t_1)=0$
and $\rho_0(q,t)=0$ for $q\in \partial \mathrm{U}$."
\end{em}

By considering $\frac{\partial I_\Sigma}{\partial \varepsilon_1}$, we deduce from this
principle \beq \label{eq14}
\partial_k\lambda(q,t) = \frac{\partial L}{\partial
w^k}\bigr(q,\frac{j(q,t)}{\rho(q,t))}\bigr), \quad (k=1,2,\,..,n) , \eeq while from
$\frac{\partial I_\Sigma}{\partial \varepsilon_2}=0$ we obtain \beq \label{eq15} \partial_t
\lambda(q,t) + \frac{j^k(q,t)}{\rho(q,t)} \frac{\partial L}{\partial w^k}
\bigr(q,\frac{j(q,t)}{\rho(q,t)}\bigr)-L=0\; , \eeq in each $\Sigma$. We note that, in the
derivation of (\ref{eq14}) and (\ref{eq15}) the condition $\rho(q,t)=0$ for
$q\in\partial\mathrm{U}$ has no role. However this condition has some implications which will
be discussed in the following.

We can solve (\ref{eq14}) with respect to $j/\rho$ and substitute the result in (\ref{eq10})
and (\ref{eq15}). To this end, we assume that our Lagrangian satisfies the condition \beq
\label{eq16} det\bigr(\frac{\partial^2 L}{\partial w^i \partial w^j}(q,w)\bigr) \neq 0 \eeq We
see that the variation of $\rho$ in the generalized action principle leads directly to consider
the classical Legendre transformation. If we introduce the Hamiltonian H \beq \label{eq17}
H(q,p) = w^k p_k - L(q,w) \,, \quad (p=(p_1,p_2,\,..,p_n)) \eeq where $w=w(q,p)$ is the
solution of the system of equations \beq \label{eq18} p_i = \frac{\partial L}{\partial w^i}
(q,w) \,,\quad  (i=1,2,\,..,n) \eeq we obtain, from (\ref{eq14}) and (\ref{eq17}), \beq
\label{eq19} \frac{j^k}{\rho} = \frac{\partial H}{\partial p_k} (q,\partial \lambda) \qquad
(\partial\lambda =
\partial_1\lambda_,..,\partial_n\lambda) \; .\eeq
Then (\ref{eq15}) becomes \beq \label{eq20} \partial_t \lambda + H(q,\partial\lambda) = 0
\qquad (q,t \in \Sigma) \eeq

In the following we will call \emph{global} any property, equation, function and so on, which
is valid or defined on the whole $\dominioRtau$, more precisely for every $q,t,\in
\dominioRtau$. Otherwise we will make use of the term \emph{local}.

In the evaluation of the global dynamical variables $\rho(q,t)$ and $j(q,t)$, to which we are
interested, we are faced with the field variables $\lambda(q,t)$, which, in principle, can be
local. This is related to the presence or absence of a set of points where $\rho(q,t)=0$. But,
a priori, we have no knowledge of this set (which we will call N). Furthermore the fields
$\lambda(q,t)$ have to satisfy, in any case, some consistency conditions, since they are
related, through (\ref{eq19}), to the variables $\rho$ and $j$. However in the equation
(\ref{eq20}) we don't have any term depending on $\rho$ (and then on the condition
$\rho(q,t)>0$). Therefore, as a result of this decoupling of $\lambda(q,t)$ from $\rho$, the
lack of knowledge of N has no influence. The consistency conditions can be solved automatically
by promoting the $\lambda(q,t)$ to a global function $S(q,t)$, thus satisfying the global
equation \beq \label{eq21}
\partial_t S(q,t)+H\bigr(q,\partial S(q,t)\bigr)=0 \qquad  (q,t \in \dominioRtau) \eeq
 From (\ref{eq19}) we deduce that $j/\rho$ can be promoted to a global variable determined
 by $S(q,t)$. Then we can write for $\rho$ the global equation
  \beq \label{eq22} \partial_t\rho(q,t) +
\partial_k \bigr( \rho(q,t)\de{H}{p_k}(q,\partial S(q,t))\bigr) =0
\quad  (q,t \in \dominioRtau) \eeq The equation (\ref{eq21}) is the well known Hamilton-Jacobi
equation \cite{bib_1,bib_2,bib_3,bib_4}. Its decoupling from (\ref{eq22}) is responsible for
the basic aspects of classical mechanics. In fact, the equation (\ref{eq22}) describes a
deterministic process. It admits, as particulare solutions, probability densities $\rho(q,t)$
having the structure
 \beq \label{eq23} \rho(q,t) = \delta(q-q(t))\; , \eeq
  where $q(t)$ satisfies the system of first order differential equations
 \beq \label{eq24} \dot{q}^k(t) =
\de{H}{p_k}\bigr(q(t),\partial S(q(t),t)\bigr) \quad (k=1,2,\,..,n) \eeq
 For this $q(t)$, one
can deduce, from (\ref{eq21}) and (\ref{eq24}), the canonical Hamilton's equations and the
Lagrange equations (\ref{eq9}). We see that the type of randomness determined by a pure
Lagrangian approach is essentially related to initial conditions in the configuration space.

From the start we have avoided deliberately references to trajectories or else to ensemble of
trajectories, with the aim to have a scheme more suitable for possible extensions. If we make
use of trajectories in the probabilistic approach, as happens if we assume initially that
$j(q,t) = \rho(q,t)v(q,t)) \quad (v\in\mathbb{R}^n)$,with $v(q,t)$ independent of $\rho$ and
regular in $\dominioRtau$, (a relation which is deduced in the previous scheme), then the
equation (\ref{eq21}) and (\ref{eq22}) can be also obtained through alternative well known
classical procedures \cite{bib_16,bib_17,bib_18}. In this respect we note also that
(\ref{eq14}) and (\ref{eq15}), with $j/\rho$ substituted by the so called \emph{"geodesic
fields" (or control fields)} have been already obtained by Carath\'{e}odory in his approach to
existence problems for the extremals of the classical action
(\ref{eq7})\cite{bib_1,bib_2,bib_19}. They have been called by him the fundamental equations of
the Calculus of Variations \cite{bib_19}. These ingredients of the Carath\'{e}odory approach
appear in a natural and unified way in the previous probabilistic scheme.

But the essential advantage of the generalized action principle is that it provides a
springboard for a suitable more general scheme. The aim is to have an unified description of
more general dynamical behaviors involving both deterministic processes given by a Lagrangian
function $L$ and processes not contemplated by this function. As a matter of fact, here we will
not be concerned with external (random) actions on our system, but with more general internal
mechanisms which govern its dynamics.

The possibility that there can be active additional mechanisms is strictly related to the
structure of our generalized action principle. In general we can implement the property that
the total probability is independent of time, through a balance equation involving $\partial_t
\rho(q,t)$ and several quantities connected with the dynamical processes taking place in our
system. It can be expressed in a form like that of a master equation. So we can naturally
extend the previous classical scheme by considering balance equations more general than the
equation (\ref{eq10}). Hence, our approach is based in general on two ingredients, a Lagrangian
function $L$ and a master-like equation, which can involve additional processes that cannot be
ascribed to this function.

However, it is useful to take into account of a further aspect of the generalized action
principle, which allows to see classical mechanics from a new perspective. It is concerned with
the non uniqueness of the Lagrangian function $L$.

In the standard variational approach, based on the trajectories, the arbitrariness of $L$ is
expressed by the addition of a total time derivative. We have a similar situation in the case
of the generalized action principle, which is strictly related to the validity of the
conservation law (\ref{eq10}). In fact, due to (\ref{eq10}), the Lagrangian $L$ and the
Lagrangian $L'$ given by \beq \label{eq25} L' = L(q,\frac j \rho) +
\frac{j^k(q,t)}{\rho(q,t)}\partial_k \chi(q,t) +
\partial_t\chi(q,t) \; , \eeq
where $\chi(q,t)$ is an arbitrary smooth function, are equivalent. They give, through the
action integral (\ref{eq12}), the same equations of motion. Therefore the choice between $L$
and $L'$ is irrelevant. However we have another class of equivalent lagrangians, involving only
our internal dynamical variables $\rho(q,t)$ and $j(q,t)$. It can be easily verified that, due
to (\ref{eq10}), also the Lagrangian $\tilde{L}$ given by \beq \label{eq26} \tilde{L}=L(q,\frac
j \rho)+ d(\rho)\frac{j^k(q,t)}{\rho(q,t)}\partial_k \rho(q,t) \; , \eeq where $d(\rho)$ is a
given function of $\rho$, is equivalent to $L$. This result is strictly related to the
condition that the variation of $\rho(q,t)$ must vanish on the boundary of $\mathrm{U}$. This
condition appears superfluous at the classical level, if we take into account of the structure
of the action functional (\ref{eq12}). However the equivalence of $L$ and $\tilde{L}$ allows us
to interpret the scheme of classical mechanics from a new point of view. In fact, if we
consider $\tilde{L}$ as our Lagrangian function, then we can say that at the classical level
there are operating also virtual internal dynamical mechanisms, generically of diffusive type.
Their manifestation is forbidden by the particular relationship between $\rho$ and $j$, given
by the balance equation (\ref{eq10}).

The fact that (\ref{eq10}) makes irrelevant the coupling between $j$ and $\partial \rho$
appearing in (\ref{eq26}), is related to a fundamental invariance property in the dynamical
behaviour of a mechanical system. Let us assume that the Lagrangian $L$ satisfies the condition
\beq \label{eq27} L(g,w) = L(g,-w) \; , \eeq as happens in general for the kinetic energy term.
Then, as is well known, we have that the law governing the evolution of our mechanical system,
given by the classical scheme, is invariant under time-reversal. Formally, the system of
equations (\ref{eq10}) and (\ref{eq20}) (with $j$ given by (\ref{eq19})) is invariant under the
(time-reversal) transformations \barr \label{eq28}
t & \rightarrow & t' = -t \nonumber \\
q^i & \rightarrow & q'^i = q^i \nonumber \\
\rho(q,t) & \rightarrow & \rho'(q,t')=\rho(q,t) \nonumber \\
\lambda(q,t) & \rightarrow & \lambda '(q,t') = -\lambda(q,t) \nonumber \\
q,t \in \Sigma & \rightarrow & q,t' \in \Sigma ' = \mathrm{U}
\times [-t_1,-t_0] \earr

Now the term $j^k\partial_k \rho$ in (\ref{eq26})) is not invariant under the transformations
(\ref{eq28})). Its ineffectiveness in the classical scheme can then be ascribed to the action
of the time-reversal invariance.

\section{\label{sec:level3}Quantum mechanics}

We consider now a particular and relevant extension of the
classical scheme.

First of all, in the framework of the generalized action principle, we take $\tilde{L}$ as the
Lagrangian function of our system, that is we take into account of the virtual internal
processes which are present at the classical level. However we don't assume that $\partial_k
j^k$ gives all the contribution to the time derivative of $\rho(q,t)$. We can interpret the
field $j(q,t)$, which through $j/\rho$ determines the value of $L$ at the point $q$, as a
quantity related in a broad sense to the transport or convective aspects of the dynamics.
Together with $\rho$, $j(q,t)$ characterizes, as before, the "state" of our mechanical system.
But now, in the balance equation, we take into account of an additional non convective
contribution to $\partial_t \rho(q,t)$ by considering, besides $j$, a current density $j_d
(q,t)$ having a given explicit dependence on $\rho$. Here we limit ourselves to a balance
equation which can be still expressed in a local form. So we generalize the equation
(\ref{eq10}) to \beq \label{eq29}
\partial_t \rho(q,t) + \partial_k j^k(q,t) + \partial_k j_d^k(q,t)
= 0 \; ,\eeq with $j_d$ a certain local function of $\rho$ \beq \label{eq30} j_d(q,t) =
i\bigr(q,\rho(q,t),\partial_l \rho(q,t),\partial_l\partial_m \rho(q,t),\, ...\bigr) \; . \eeq

As a matter of fact, through the current $j_d$ in the balance equation we consider a real
additional mechanism in the dynamical evolution, which is, in a broad sense, of diffusive type.
However it makes real also the processes contemplated by $\tilde{L}$, which are virtual when
$j_d =0$. Starting from $\tilde{L}$ and the extended balance equation (\ref{eq29}), the action
integral in the generalized action principle becomes \barr \label{eq31}
 \tilde{I}_\Sigma[j,\rho,\lambda] &=&\int_\Sigma
dtdq\biggr\{\bigr[L(q,\frac{j}{\rho})+d(\rho)\frac{j^k}{\rho}\partial_k\rho
\bigr]\rho(q,t) + \nonumber \\ & +&\lambda \bigr(\partial_t \rho
+\partial_k j^k + \partial_k i^k(q,\rho,\partial_l \rho,..)\bigr)
\biggr\} \earr From (\ref{eq31}) we deduce the generalization of
(\ref{eq14}) and (\ref{eq15}), that is \beq \label{eq32}
\partial_k \lambda - d(\rho)\partial_k\rho=\de{L}{w^k}(q,\frac j
\rho)\qquad (k=1,2,\,..,n)\; , \eeq \\ \barr \label{eq33} \partial_t \lambda & + &
\frac{j^k}{\rho} \de{L}{w^k}(q, \frac{j}{\rho}) - L + d(\rho)\partial_k j^k + (\partial_k
\lambda) \de{i^k}{\rho} - \nonumber \\ & - & \partial_k \bigr( (\partial_l \lambda)
\de{i^l}{(\partial_k \rho)}\bigr) + \partial_k \partial_l \bigr( (\partial_m \lambda)
\frac{\partial i^m}{\partial(\partial_k
\partial_l\rho)}\bigr) - ... = 0\; , \nonumber \\ \earr in every
$\Sigma$.

By introducing the Hamiltonian $H$, we obtain the system of the two coupled equations \barr
\label{eq34} \partial_t \lambda & + & H(q,\partial \lambda - d(\rho)\partial \rho) + \nonumber
\\ &+& d(\rho)\partial_k \bigr( \rho \de{H}{p_k}(q,\partial \lambda - d(\rho)\partial \rho)
\bigr) + \nonumber \\ &+& (\partial_k \lambda) \de{i^k}{\rho} -
\partial_k \bigr( (\partial_l \lambda) \de{i^l}{(\partial_k \rho)}\bigr) + \nonumber \\ &+& \partial_k \partial_l \bigr(
(\partial_m \lambda) \frac{\partial i^m}{\partial(\partial_k
\partial_l\rho)}\bigr) - ... = 0 \earr \beq \label{eq35} \partial_t \rho + \partial_k \biggr( \rho
\de{H}{p_k} (q,\partial \lambda - d(\rho)\partial \rho)) \biggr) +
\partial_k i^k=0 \; , \eeq in every $\Sigma$.

In general, in the system of equations (\ref{eq34}) and (\ref{eq35}), we lose time reversal
invariance. However we impose to our extension of the classical scheme the explicit condition
that the invariance under the transformation (\ref{eq28}) is preserved, notwithstanding the
presence of the current $i(q,\rho,\partial_l \rho,\,..)$ and of $d(\rho)\partial \rho$. As a
matter of fact we have two types of diffusive processes, which can have a competitive or
antagonistic role. As we will see, the requirement of time-reversal invariance selects an
important class of mechanical systems. Moreover, it allows to deduce completely, from the
knowledge of $\tilde{L}$, the specific structure of $j_d$.

A condition for the invariance under the transformations (\ref{eq28}) is that $d(\rho)$ must be
necessarily different from zero when $j_d \neq 0$, as can be easily seen from the equations
(\ref{eq35}). Moreover, by considering (\ref{eq34}), we deduce from (\ref{eq28}) that this
invariance requires \barr \label{eq36} \frac{1}{2}&\biggr[&H(q, \partial \lambda -
d(\rho)\partial\rho) - H(q, -\partial \lambda - d(\rho)\partial\rho)\;\biggr] + \nonumber
\\ &+& d(\rho)\partial_i\biggr[ \frac{\rho}{2}
\biggr(\de{H}{p_i}(q,
\partial \lambda - d(\rho)\partial\rho\bigr) - \de{H}{p_i}\bigr(q, -\partial \lambda - d(\rho)\partial\rho\biggr)\biggr]\nonumber \\ &+& (\partial_k \lambda)\de{i^k}{\rho} - \partial
_k\biggr( (\partial_l\lambda)\de{i^l}{\partial_k \rho}\biggr)+
\nonumber \\ &+& \partial_k \partial_l \biggr( (\partial_m
\lambda) \frac{\partial i^m}{\partial(\partial_k
\partial_l\rho)}\biggr) - ... = 0\; .\earr

From the structure of the equation (\ref{eq36}) and the requirement that it must be satisfied
identically by $\lambda$ and $\rho$, we deduce that $H(q,
\partial \lambda - d(\rho)\partial\rho) - H(q, -\partial \lambda - d(\rho)\partial\rho)$ must
be linear in $\partial\lambda$, or, in other words, that $H(q,p)$ must be quadratic in $p$.
 So, the request that the system (\ref{eq34}),(\ref{eq35}) be invariant under the time-reversal
transformations (\ref{eq28}), forces us to limit ourselves to natural \cite{bib_20} Lagrangian
systems, that is to mechanical systems described by a Lagrangian of the type \beq \label{eq37}
L = L (q,w) = \frac{1}{2} m_{ij}(q)w^i w^j - V(q)\,, \eeq where $m_{ij}(q) ( = m_{ji}(q))$ is a
positive definite mass matrix.

From (\ref{eq36}) we deduce also \beq \frac{\partial
i^m}{\partial(\partial_k\partial_l\rho)}=\frac{\partial i^m}{\partial(\partial_k
\partial_l\partial_j\rho)}= ... = 0 \; ,\nonumber \eeq so that \beq
\label{eq38} j_d = i(q,\rho,\partial \rho). \eeq Hereafter we will fix our attention on the
Lagrangian (\ref{eq37}). We have \beq \label{eq39} H(q,p) = \frac{1}{2} m^{ij}(q)p_i p_j + V(q)
\; ,\eeq where $m^{ij}$ is the inverse matrix of $m_{ij}$. The condition (\ref{eq36}) of
time-reversal invariance becomes \barr \label{eq40} \partial_k \lambda \biggr[ \rho
d(\rho)\partial_j m^{ij} + \de{i^k}{\rho}-\dpr j\biggr(\de{i^k}{(\partial_j
\rho)}\biggr)\biggr] + \nonumber \\ + \,\partial_k \partial_j \lambda \biggr[ \rho
d(\rho)m^{kj} -\de{i^k}{(\partial_j \rho)}\biggr] = 0 \; .\earr

In order that (\ref{eq40}) be satisfied identically by $\lambda$ and $\rho$, we must have a
relationship between $d(\rho)$ and the current $i(q,\rho,\partial \rho)$. It is given by \beq
\label{eq41} i^k(q,\rho,\partial \rho)=m^{kj}(q)\rho\, d(\rho)\partial_j \rho+ \tilde{i}^k(q)
\qquad (k=1,2,\,..,n) \eeq

It can be easily verified that (\ref{eq41}) makes also the equation (\ref{eq35}) time-reversal
invariant, provided $\partial_k \tilde{i}^k(q)=0$. So $\tilde{i}^k(q)$ is irrelevant and can be
completely neglected. We conclude that $i^k$ must be a diffusion current of the type \beq
\label{eq42} i^k = -D^{kj}(q,\rho) \partial_j \rho\; , \eeq with a matrix diffusion coefficient
determined by the mass matrix and by the "coupling" function $d(\rho)$ (so that in general it
depends on $\rho$), i. e. \beq \label{eq43} D^{kj}(q,\rho) = -\rho d(\rho) m^{kj}(q)\; . \eeq

So, we have worked out a formulation of a class of dynamical behaviors, which generalizes the
classical scheme, but preserves in any case the property of time-reversal invariance. This
class is characterized by the action functional (\ref{eq31}), the Lagrangian (\ref{eq37}) and
the relationship (\ref{eq41}) (with $\tilde{i}^k=0$). The law governing these dynamical
behaviors follows from (\ref{eq34}) and (\ref{eq35}). We have \barr \label{eq44} \partial_t
\lambda &+& \frac{1}{2}m^{ij}\partial_i\lambda\partial_j\lambda + V +
\frac{1}{2}m^{ij}\bigr(\rho d^2(\rho)\bigr)' \partial_i \rho
\partial_j \rho \nonumber \\ &-&
\partial_i\bigr(m^{ij}\rho d^2(\rho)\partial_j\rho\bigr)=0 \; ,\earr
\beq \label{eq45} \partial_t \rho + \partial_i (\rho\: m^{ij}
\partial_j \lambda)=0\; , \eeq
in every $\Sigma$ where $\rho > 0$ $\bigr( (\,)' =
\frac{d}{d\rho}() \bigr)$.

It is useful to note that the system (\ref{eq44}),(\ref{eq45}) has an Hamiltonian structure,
with $\rho(q,t)$ and $\lambda(q,t)$ playing the role of conjugate fields variables. From
(\ref{eq31}) and (\ref{eq32}) it follows easily that (\ref{eq44}) and (\ref{eq45}) can be
deduced directly from the standard variational principle applied to the action functional \beq
\label{eq46} A_\Sigma [\lambda,\rho] = \int_\Sigma dt dq \biggr[-\rho\partial_t\lambda -
\mathcal{H}_e\biggr]\; , \eeq with the effective Hamiltonian density $\mathcal{H}_e$ given by
\barr \label{eq47} \mathcal{H}_e(\rho,\partial\rho,\lambda,\partial\lambda) &=& +\rho
\biggr[\frac{1}{2}m^{kl}\partial_k\lambda\partial_l\lambda + V\biggr]+\nonumber \\
&+&\frac{1}{2}\rho d^2(\rho)m^{kl}\partial_k\rho\partial_l\rho\; . \earr (The variation of
$\lambda$ must vanish on the boundary of $\Sigma$). A particular case of \rif{47} has been
previously postulated in the F\'{e}nyes theory \cite{bib_21}.

Now, our problem is to determine the global field variables $\rho(q,t)$ and $j(q,t)$, having at
our disposal the two coupled local equations (\ref{eq44}),(\ref{eq45}) and the local
relationship (\ref{eq32}). From this we deduce, as before, that the local (in principle) fields
$\lambda(q,t)$ must satisfy some consistency conditions. In this respect we observe that if
$\rho d^2 (\rho)$ is a smooth function of $\rho$ regular at $\rho=0$, there appears no
effective restriction on the set of points in $\dominioRtau$ in which the equations
(\ref{eq44}) and (\ref{eq45}) can be considered valid. In this case, the Hamiltonian
$\mathcal{H}_e$ is a smooth function in every domain of $\dominioRtau$. As a consequence, in
this case, the $\lambda(q,t)$ can be promoted to a global field variable $S(q,t)$, as happens
in the classical scheme. But now the equation for $S(q,t)$ involves also $\rho(q,t)$. Such a
situation will be called the extended classical domain. We can make a further step and go
beyond the extended classical domain, by considering the situation in which $\rho d^2 (\rho)$
is singular at $\rho=0$. In this case the above straightforward globalization does not apply.
However we can ask if this singularity can be absorbed in a canonical transformation of the
field variables $\rho(q,t)$ and $\lambda(q,t)$, allowing, in terms of the new variables, a
straight-forward global description of dynamical behaviors not contemplated by the extended
classical domain.

By fixing the attention on a set $\Sigma$, let us express $\rho$ and $\lambda$ in terms of two
new variables \beq \label{eq48} \rho = P(u,v) \qquad \lambda=\Lambda(u,v)\; .\eeq We will
assume that the transformation defined in (\ref{eq48}) is canonical, i. e. it has the property
that there exists a function $F(u,v)$ such that \beq \label{eq49} P d\Lambda = u du + dF \eeq
By considering $u$ and $v$ as fields, $u = u(q,t)$ and $v = v(q,t)$, the system
(\ref{eq44}),(\ref{eq45}) will be transformed in a new system involving $u = u(q,t)$ and $v =
v(q,t)$, having still a canonical structure.

A necessary and sufficient condition that (\ref{eq48}) be canonical is \beq \label{eq50}
\de{(P,\Lambda)}{(u,v)} = \de{P}{u} \de{\Lambda}{v} - \de{P}{v} \de{\Lambda}{u} = +1 \eeq

Now, suppose that $\rho d^2(\rho)$ is singular at $\rho=0$. Since we consider $\rho$,$j$ and
$j_d$, which appear in the global balance equation (\ref{eq29}), as global field variables, we
deduce from (\ref{eq41}) that we must have in any case that $\rho d(\rho)$ is regular at
$\rho=0$. This means that $\rho d^2(\rho)$ must have a pole of first order at $\rho=0$. So we
can write \beq \label{eq51} \rho d^2(\rho)=\frac{(a/2)^2}{\rho} + g(\rho) \qquad (a>0)\,,\eeq
where $a$ is a constant and $g(\rho)$ is a smooth function of $\rho$, regular at $\rho=0$.

By considering the canonical transformation (\ref{eq48}), we have \beq \label{eq52} -\rho
\partial_t \lambda - \mathcal{H}_e = -u(q,t) \partial_t v(q,t) - {\mathcal{H}'}_e(u,\partial u,
v, \partial v) + \partial_t F(u,v) \; ,\eeq where, according to (\ref{eq51}), the transformed
effective Hamiltonian density is given by \barr \label{eq53} {\mathcal{H}'}_e &=& \frac{1}{2}
m^{kl}\biggr\{ \biggr[P(\de{\Lambda}{u})^2 +
\frac{(a/2)^2}{P}\biggr(\de{P}{u}\biggr)^2\biggr]\dpr{k}u\dpr{l}u - \nonumber \\ &-&
\biggr[P\biggr(\de{\Lambda}{v}\biggr)^2 +
\frac{(a/2)^2}{P}\biggr(\de{P}{v}\biggr)^2\biggr]\dpr{k}v\dpr{l}v - \nonumber \\ &-& 2
\biggr[P\biggr(\de{\Lambda}{u}\biggr)\biggr(\de{\Lambda}{v}\biggr) +
\frac{(a/2)^2}{P}\biggr(\de{P}{u}\biggr)\biggr(\de{P}{v}\biggr)\biggr]\dpr{k}u\dpr{l}v
\biggr\}\nonumber \\ &+& \frac{1}{2}m^{kl}g(P)\dpr k P \dpr l P + PV \; .\earr

Our aim is to eliminate, through an appropriate choice of the canonical transformation, the
singular dependence of ${\mathcal{H}'}_e$ on $P$, at $P=0$. If this happens, the field
variables involved in this transformation can be promoted straight-forwardly to global
variables. To this end, on the basis of the structure of the coefficients of the quadratic form
appearing in (\ref{eq53}) and of the condition (\ref{eq50}), we consider a transformation such
that \beq \label{eq54} \frac a 2 \de P u = P \de \Lambda v \qquad \frac a 2 \de P v = -P \de
\Lambda u \eeq

If (\ref{eq50}) and (\ref{eq54}) are satisfied, we have \beq
P\biggr(\de{\Lambda}{u}\biggr)\biggr(\de{\Lambda}{v}\biggr) +
\frac{({a/2})^2}{P}\biggr(\de{P}{u}\biggr)\biggr(\de{P}{v}\biggr) =0\; , \nonumber \eeq
 \barr \label{eq55} P\biggr(\de{\Lambda}{u}\biggr)^2 + \frac{(a/2)^2}{P}\biggr(\de{P}{u}\biggr)^2 &=&
P\biggr(\de{\Lambda}{v}\biggr)^2 +
\frac{(a/2)^2}{P}\biggr(\de{P}{v}\biggr)^2 \nonumber \\
&=& \frac a 2 \; .\earr
 So the singular dependence on $P$, at $P=0$, disappears. Now, let us determine the above canonical
 transformation. From  (\ref{eq54}) it follows that $\log P$ and $\Lambda\bigr/(a/2)$ must be the
 real and the imaginary part of an analytic
 function $f(z)$ \beq \label{eq56} \log P + \emph{i} \frac{\Lambda}{a/2} = f(z)\,,
 \eeq with $z = u + \emph{i} v$. Furthermore, due to (\ref{eq50}),
 we have \beq \label{eq57} \bigr|f'(z)\bigr|^2 = \biggr(\de{\log
 P}{u}\biggr)^2+\biggr(\de{}{u} \frac{\Lambda}{a/2}\biggr)^2 =
 \frac{1}{P \cdot a/2} \eeq Then, by making use of (\ref{eq56}),
 we deduce that $f(z§)$ satisfies the equation \beq \label{eq58} \biggr| \dtot{}{z} e^{f(z)/2}\biggr|^2 =
 \frac{1}{2a} \eeq Therefore \beq \label{eq59} e^{\frac{f(z)}{2}} = \biggr(\frac{1}{2a}\biggr)^{\frac 1 2} e^{\emph{i}
 \varphi}(z+c_1+\emph{i} c_2) \,,\eeq where $\varphi$,$c_1$,$c_2$ are real constant.

 In conclusion, the canonical transformation which satisfies
 \rif{54} is given by \barr \label{eq60} P(u,v) =
 \frac{(u+c_1)^2+(v+c_2)^2}{2a} \nonumber \\ \frac{\Lambda(u,v)}{a} =
 arg( u + c_1 + \emph{i}(v+c_2)) + \varphi \earr
 (In $\Sigma$ we can select a determination of arg). So, as a consequence of \rif{51}, in this case
the probability density is related to the squared modulus of a complex amplitude, while
$\Lambda$ is connected with the phase of this amplitude.
 Such a situation will be called in general the extended quantum domain.
 The constant $c_1$ and $c_2$ can be absorbed in a further
 canonical transformation: $u'=u+c_1 \; , \; v'=v+c_2$.
 Furthermore it is useful to use the scaled field variables
 \beq \psi_1 = \frac{u'}{(2a)^\frac 1 2} \qquad \psi_2 = \frac{v'}{(2a)^\frac 1
 2} \,,\nonumber \eeq so that \rif{60} becomes
 \beq \label{eq61} P(\psi_1,\psi_2) = \psi_1^2 + \psi_2^2 \;,\qquad \frac{\Lambda(\psi_1,\psi_2)}{a} =
 \arctan{\frac{\psi_1}{\psi_2}} + \varphi \eeq
 By considering the effective lagrangian \beq \label{eq62}
\mathcal{L}_e = -\rho \dpr t \lambda - \mathcal{H}_e\,,\eeq
 which gives the system \rif{45},\rif{45}, we deduce, from
 \rif{52},\rif{53},\rif{54},\rif{55} and \rif{61}, that
 $\mathcal{L}_e$ is given also by \barr \label{eq63} \mathcal{L}_e =
 &-& a\bigr(\psi_1 \dpr t \psi_2-\psi_2\dpr t \psi_1\bigr)+ \nonumber \\&+&\frac 1 2
 a^2 m^{kl}\bigr(\dpr k \psi_1 \dpr l \psi_1 + \dpr k \psi_2 \dpr l \psi_2\bigr)- \nonumber \\
 &-& V(q)(\psi_1^2 + \psi_2^2) - \nonumber \\&-& 2g(\psi_1^2 + \psi_2^2)m^{kl}\bigr(\psi_1
 \dpr k \psi_1+\psi_2\dpr k \psi_2\bigr)\cdot
\nonumber \\ && \cdot \bigr(\psi_1 \dpr l \psi_1+\psi_2\dpr l \psi_2\bigr)\,, \nonumber \\
 \earr where we have neglected a term involving a total time derivative.

 Although we have derived \rif{63} in a set $\Sigma$, we see that all the coefficients of
the quadratic form involving $\dpr k \psi_1$,$\dpr l \psi_2$ are regular functions, even at
$\rho = \psi_1^2 + \psi_2^2 = 0$. So there is no effective restriction to consider
$\mathcal{L}_e$, as given by \rif{63}, on the whole $\dominioRtau$and to promote
straight-forwardly $\psi_1(q,t)$ and $\psi_2(q,t)$ to global field variables. From \rif{63} we
can deduce a system of two equations for $\psi_1$ and $\psi_2$, which globalizes the system
\rif{44},\rif{45}.

 In this respect, it is useful to consider a further canonical transformation, by introducing a
complex field $\psi$ and its conjugate
 \beq \label{eq64} \psi = \psi_1 + \emph{i} \psi_2 \qquad \psi^\ast
 = \psi_1 - \emph{i} \psi_2 \eeq
 In terms of the field variables $\psi$ and $\psi^\ast$, the
 Lagrangian \rif{63} becomes
 \barr \label{eq65} \mathcal{L}_e =
 &+& \emph{i} \frac a 2 \bigr(\psi^\ast \dpr t \psi-\psi\dpr t \psi^\ast\bigr)- \nonumber \\&-&\frac 1 2
 a^2 m^{kl} \dpr k \psi^\ast \dpr l \psi - V \psi^\ast \psi \nonumber \\ &-& \frac{g(\psi^\ast\psi)}{2}m^{kl}\bigr(\psi^\ast \dpr k \psi+\psi\dpr k \psi^\ast\bigr)\bigr(\psi^\ast \dpr l \psi+\psi\dpr l \psi^\ast\bigr) \nonumber \\
 \earr
 From \rif{65} we deduce a system of two equations for $\psi$ and
 $\psi^\ast$. One of these equations is the complex conjugate of
 the other. In general, there is a coupling between the two
 equations. The usefulness of the canonical fields $\psi$ and
 $\psi^\ast$, stems from the fact that they allow to determine a
 single and relevant circumstance in which we obtain two decoupled
 equations. This happens when \beq \label{eq66} g(\psi^\ast \psi)
 = 0 \eeq

 In this case, we have a situation analogous (but more
 symmetrical) to that of the classical domain, in which a
 canonical field is decoupled from the other. In the classical
 domain the decoupling is responsible for the existence of
 deterministic trajectories. The decoupling obtained when \rif{66}
 is satisfied, is responsible for the linearity of the resulting
 equation for $\psi$ (or $\psi^\ast$). In this case we speak of
 the quantum domain. When \rif{66} is satisfied, we deduce from
 \rif{65} \beq \label{eq67} \emph{i} \,a \dpr t \psi(q,t) = - \frac{a^2}{2} \dpr i \biggr( m^{ij}(q)\dpr j \psi(q,t)\biggr) +
 V(q)\psi(q,t) \eeq

 For
 \beq \label{eq68} a = \hbar\,,\eeq \rif{67} is the Schr\"{o}dinger
 equation which governs the quantum mechanics of a natural
 mechanical system. If $g(\psi^\ast \psi) \neq 0$, we obtain a non linear Schr\"{o}dinger equation.

 It is useful to stress that the need for the field variables like
 $\psi_1(q,t)$ and $\psi_2(q,t)$ is related to the fact that our
 problem is to determine the global field variables $\rho(q,t)$
 and $j(q,t)$, while at an intermediate step we have to introduce
 a local (in principle) field variable, that is $\lambda(q,t)$. We
 are then faced with the solution of some matching conditions,
 which among other things would require a knowledge (the set N)
 not given a priori. These conditions are automatically solved by
 considering new field variables for which there is no
 obstacle to be extended globally in straight-forward way. Without
 any condition we could not say that system of local equations
 \rif{44} and \rif{45} is equivalent, for $g(\psi^\ast \psi) = 0$,
 to the Schr\"{o}dinger equation, which is valid globally. As a
 matter of fact \rif{44} and \rif{45} is a system of generalized
 Madelung hydrodynamic equations \cite{bib_15,bib_18}. The
 inequivalence between the Schr\"{o}dinger equation and the
 Madelung equations has been already discussed by T.C. Wallstrom
 \cite{bib_22}. However it is significant that, when \rif{66} is
satisfied, the system of local equations \rif{44} and \rif{45} can be interpreted in terms of
classical Brownian trajectories, as shown in the Nelson's stochastic mechanics approach
\cite{bib_16,bib_17,bib_23,bib_24}.

 Our previous analysis provides, as a particular relevant case,
a scheme for a quantization procedure. In this respect we note that, in the framework of the
canonical quantization rules, we have in general a problem with the ordering of the product of
non commuting operators, as happens when an Hamiltonian of the type \rif{39} is generally
considered. Our scheme, through the equation \rif{67}, allows to give an answer to this
problem.
 Furthermore we remark that the consideration of the Lagrangian
$\mathcal{L}_e$ $\bigr($\rif{63} or \rif{65}$\bigr)$ is also useful from another point of view.
By applying to $\mathcal{L}_e$ the No\"{e}ther theorem \cite{bib_3}, we can select the relevant
conserved physical quantities, associated to the symmetry properties of our mechanical system.
When \rif{66} is satisfied, these quantities are in correspondence with the expectation values
of the operators of the standard quantum theory \cite{bib_18}.

From the theoretical point of view, our approach predicts different dynamical behaviours,
according to the structure of $d(\rho)$. This structure cannot be given a priori, without
further considerations. In principle it must be argued on physical grounds.

 The quantization procedure afforded by the previous scheme
requires that the mechanical system be natural, a property not necessarily contemplated by the
canonical quantization rules. We could also develope a "quantum" theory of a non natural
mechanical system, on the basis of the action \rif{31} or the system \rif{34},\rif{35}. The
resulting "quantum" dynamical behavior would break the time reversal invariance of the
classical level. We note that a mechanical system with relativistic kinematics is non natural.
In this case we are also faced with an "homogeneous" problem in the calculus of variations
\cite{bib_2,bib_19}.

In the case of "homogeneous" problems, or problems in parametric form, the condition \rif{16}
is not satisfied. In order to treat these cases (singular lagrangians), within our approach, we
need a further elaboration as done in the calculus of variations or in the Dirac approach to
constrained hamiltonian systems \cite{Hanson}. The limiting procedure of H. J. Rothe
\cite{Rothe} could be also useful. These aspects will be investigated elsewhere.

 We conclude this section with a provisional discussion of another extension of the classical scheme,
which can be formulated in the framework of our approach.

 Suppose to consider the time evolution in the interval $[0,\tau]$
of a dynamical system whose configurations are described by a discrete variable assuming a
finite number N of values. These values will be labelled by a discrete index running from one
to N. At each instant $t \in [0,\tau]$ we consider the configuration of the system a discrete
random variable described by a set of probabilities $p_\alpha(t) \quad (\alpha = 1,2,..,N)$.

 As a first step, in order to take into account of the fact that
 \beq \sum_{\alpha=1}^N p_\alpha(t) \nonumber \eeq
 does not depend on $t$, we introduce a quite general balance
 equation in the form \beq \label{eq69} \dot{p_\alpha}(t) +
 \sum_{\beta=1}^N
 \bigr(\gamma_{\alpha\beta}(t)-\gamma_{\beta\alpha}(t)\bigr)=0
 \qquad (\alpha=1,2,..,N)\eeq
The non local and discrete equation \rif{69} takes the place of \rif{10}. We plan to determine
variationally $p_\alpha(t)$ and $\gamma_{\alpha\beta}(t)$. In this respect, here we will assume
that the functions $p_\alpha(t)$, for each $\alpha$, are such that they have at most a finite
number of zeros in the interval $[0,\tau]$. As a second step we introduce a "Lagrangian"
$\mathcal{L}(\xi)$, a smooth function of a real variable $\xi$, and a real symmetric matrix
$\mathrm{U}_{\alpha\beta}$, time independent, which will be connected with the extension of the
local substitution \rif{11}. We will assume that \beq \label{eq70} \mathcal{L}(-\xi) =
\mathcal{L}(\xi) \eeq and \beq \label{eq71} \mathrm{U}_{\alpha\beta} = \mathrm{U}_{\beta\alpha}
\neq 0 \qquad \forall \alpha,\beta \eeq Suppose now that $p_\alpha(t)$ and
$\gamma_{\beta\alpha}$ give an actual temporal evolution of our system and let us consider the
family of subset $\Gamma$ of $[0,\tau]$, formed by finite unions of disjoint closed internals
in which $p_\alpha(t)>0$, for every $\alpha$. As a final step, in each $\Gamma$ we extend
\rif{11} by the "non local" substitution \beq \label{eq72} \xi \; \rightarrow \;
\frac{\gamma_{\alpha\beta}(t)}{p_\alpha^\frac 1 2(t)\mathrm{U}_{\alpha\beta}p_\beta^\frac 1
2(t)}\,. \eeq Then we replace the action functional \rif{12} by \barr \label{eq73}
&&I_\Gamma[\gamma_{\alpha\beta},p_\alpha,\lambda_\alpha] =\nonumber
\\ &&= \int_\Gamma
dt\sum_{\alpha=1}^N\biggr\{\sum_{\beta=1}^N\mathcal{L}\biggr(\frac{\gamma_{\alpha\beta}(t)}{p_\alpha^\frac
1 2(t)\mathrm{U}_{\alpha\beta}p_\beta^\frac 1
2(t)}\biggr)p_\alpha^\frac 1
2(t)\mathrm{U}_{\alpha\beta}p_\beta^\frac 1 2(t) - \nonumber \\
&& - \lambda_\alpha (t) \biggr(\dot{p}_\alpha(t)+
\sum_{\beta=1}^N\bigr(\gamma_{\alpha\beta}(t)-\gamma_{\beta\alpha}(t)\bigr)\biggr)\biggr\}
\earr
 By applying straight-forwardly the generalized action principle
we deduce from \rif{73} \beq \label{eq74} \lambda_\alpha(t) - \lambda_\beta(t) =
\dtot{\mathcal{L}}{\xi} \biggr(\frac{\gamma_{\alpha\beta}(t)}{p_\alpha^\frac 1
2(t)\mathrm{U}_{\alpha\beta}p_\beta^\frac 1 2(t)}\biggr) \quad (\alpha,\beta = 1,..N;
t\in\Gamma) \eeq By introducing again a Legendre transformation \beq \label{eq75}
\mathcal{H}(\eta) = \eta\xi - \mathcal{L}(\xi)\,, \eeq we can write \barr \label{eq76}
\gamma_{\alpha\beta}(t) = p_\alpha^\frac 1 2(t)\mathrm{U}_{\alpha\beta}p_\beta^\frac 1
2(t)\dtot{\mathcal{H}}{\eta}(\lambda_\alpha(t)-\lambda_\beta(t))&& \nonumber \\ (\alpha,\beta =
1,..N; t\in\Gamma) \earr

Then we obtain the system of equations: \beq \label{eq77} \dot{\lambda}_\alpha -
\sum_{\beta=1}^N \mathrm{U}_{\alpha\beta} \frac{p_\beta^\frac 1 2}{p_\alpha^\frac 1
2}\mathcal{H}(\lambda_\alpha-\lambda_\beta)=0\eeq \barr \label{eq78} \dot{p}_\alpha + 2
\sum_{\beta=1}^N  p_\alpha^\frac 1 2(t)\mathrm{U}_{\alpha\beta}p_\beta^\frac 1
2(t)\dtot{\mathcal{H}}{\eta}(\lambda_\alpha-\lambda_\beta) = 0 \nonumber \\ (\alpha,\beta =
1,..N; t\in\Gamma)\,, \earr where we have made use of \rif{70},\rif{71} and \rif{76}.

The system \rif{77},\rif{78} defines an Hamiltonian structure. It can be deduced directly from
the action functional \barr \label{eq79} A_\Gamma[\lambda_\alpha,p_\alpha]
&=&\int_\Gamma\biggr[-\sum_{\alpha=1}^N p_\alpha(t)\dot{\lambda}_\alpha(t)+\nonumber
\\ &+&\sum_{\alpha,\beta=1}^N p_\alpha^\frac 1
2(t)\mathrm{U}_{\alpha\beta}p_\beta^\frac 1 2(t)
\mathcal{H}(\lambda_\alpha-\lambda_\beta)\biggr]  \earr

We see that the local character of the system \rif{77},\rif{78} is determined by the
non-analytic square root dependence on $p_\alpha(t)$ of the action functional \rif{79}. Then we
can obtain an explicit description of a global dynamical behavior, if we can absorb this
dependence in a canonical transformation involving variables which can be extended globally
without any condition. However, in order to carry out this plan, we have to select
appropriately the structure of $\mathcal{H}(\eta)$. As a matter of fact our goal can be
realized quite easily, by choosing as our "Hamiltonian" \beq \label{eq80} \mathcal{H}(\eta)= b
\, \cos{\frac \eta a} \qquad (a >0)\,,\eeq where $a$ and $b$ are real constants. We have then
immediately the appropriate canonical transformation, which, written in the standard form
involving complex quantities, is given by \beq \label{eq81} \psi_\alpha(t) = p_\alpha^{\frac 1
2}\; e^{\emph{i} \frac{\lambda_\alpha(t)}{a}} \eeq

The action functional \rif{79} becomes \barr \label{eq82} A_\Gamma[\lambda_\alpha,p_\alpha] =
A_\Gamma[\psi_\alpha] &=&  \int_\Gamma \biggr[\emph{i} \frac a 2 \sum_{\alpha=1}^N\bigr(
\psi_\alpha^\ast \dot{\psi_\alpha} - \psi_\alpha\dot{\psi_\alpha}\bigr)+\nonumber \\ &+&
\sum_{\alpha,\beta=1}^N\psi_\alpha^\ast\,b\,\mathrm{U}_{\alpha\beta}\psi_\beta\biggr] \earr

In terms of the variables $\psi_\alpha(t)$ we obtain straight-forwardly a global description of
the dynamical behavior. As a matter of fact the $\psi_\alpha(t)$ which are solutions of the
evolution equations are analytic functions of $\Gamma$, so that our assumption on the zeros of
$p_\alpha(t)$ is satisfied.

We consider two generalizations of the previous result. We can introduce a possible constant
shift in the stationary point of $\gamma_{\alpha\beta}(t)$, by adding in the integrand of
equation \rif{73} a term linear in $\gamma_{\alpha\beta}(t)$ \beq \label{eq83} -
\sum_{\alpha,\beta=1}^N \theta_{\alpha\beta}\gamma_{\alpha\beta}(t)\,,\eeq where the given real
matrix $\theta_{\alpha\beta}$ is antisymmetric $(\theta_{\alpha\beta} =
-\theta_{\beta\alpha})$. On the other hand, through an appropriate limiting procedure, we can
consider also the case in which some elements of the matrix $\mathrm{U}_{\alpha\beta}$ are
equal to zero. Then we obtain the final global effective action functional \beq \label{eq84}
A_\Gamma[\psi_\alpha] = \int_0^\tau \biggr[\emph{i} \frac a 2 \sum_{\alpha=1}^N\bigr(
\psi_\alpha^\ast \dot{\psi_\alpha} - \psi_\alpha\dot{\psi_\alpha}^\ast\bigr)-
\sum_{\alpha,\beta=1}^N\psi_\alpha^\ast\,h_{\alpha\beta}\psi_\beta\biggr]\,,\eeq with a generic
hermitian matrix $h_{\alpha\beta}$ given by \beq \label{eq85} h_{\alpha\beta} = -b\,
\mathrm{U}_{\alpha\beta}\,e^{-\emph{i} \,\frac{\theta_{\alpha\beta}}{a}} \eeq

From \rif{84} we deduce the standard discrete Schr\"{o}dinger equation \beq \label{eq86}
\emph{i} \hbar\, \dot{\psi}_\alpha(t) = \sum_{\beta=1}^N h_{\alpha\beta}\psi_\beta(t) \qquad
(\alpha=1,2,..\,,N)\,, \eeq which governs the quantum dynamics of a discrete random variable,
like the spin.

We note finally that it possible to consider also the situation in which we have both
continuous and discrete random variables.

\section{\label{sec:level4}Probabilistic variational approach to classical field theory}

Let us come back to our physical system described classically by a set of real fields $q^i(x)$
and the Lagrangian density \rif{1}. By following the theoretical scheme of
Sect.\ref{sec:level2}, we adopt at the start a probabilistic point of view and, at the same
time, we plan to generalize the standard variational principle based on the action \rif{2}.

At first sight, the immersion of the deterministic classical field theory in a probabilistic
scheme could be realized immediately if a field is considered as a m.s.i. As a matter of fact
the mechanical transcription allows to obtain an Hamilton-Jacobi equation. In this case, we
have a functional equation, involving a functional $S\bigr([q(\vec{x}),t]\bigr)$, for each $t$,
of the configurations $q(\vec{x})$ of our system of fields $\bigr(\vec{x}=(x^1,x^2,x^3)\bigr)$.
However, we meet with difficulties if we want to place at the side of
$S\bigr([q(\vec{x}),t]\bigr)$, a functional probability density
$\rho\bigr([q(\vec{x}),t]\bigr)$. There is no Lebesgue measure in an infinite dimensional space
and there are problems with a functional divergence theorem. So we are faced with a
probabilistic scheme which is already ill-defined at the classical level, before the transition
to the quantum theory. Furthermore, space and time are treated in an asymmetric way. On the
other hand, as will be discussed in the following, it is possible to embed deterministic
classical field theory in a simple and well-defined probabilistic scheme in which the above
difficulties are avoided.

At each space-time point $x$, we treat the values $q=(q^1,q^2,..\,,q^n)$ of the set of our
fields as random variables described by a probability density $\rho(q,x)$. We have now that the
total probability \beq \int_{\mathbb{R}^n}\rho(q,x)dq \nonumber \eeq does not depend on the
space-time point $x$. We assume again that this independence can be expressed in a local form
\beq \label{eq87} \dpr{\mu}\rho(q,x)+\dpr k j_\mu^k(q,x) = 0 \qquad (\mu=0,1,2,3) \eeq So,
together with $\rho(q,x)$, we have now four probability current densities
$j_\mu(q,x)=\bigr(j_\mu^1(q,x),..\,,j_\mu^n(q,x)\bigr)$ through which the "state" of the system
is characterized. In terms of $\rho(q,x)$ and $j_\mu(q,x)$ we can extend in a natural way to
fields the generalized Hamilton's principle of sect.\ref{sec:level2}.

It is formally convenient to take into account of the property that the Lagrangian density
\rif{1} is a function of $q$ and of $4n$ independent variables $w_\mu =
(w_\mu^1,..\,,w_\mu^n)$, \beq \label{eq88} \mathcal{L}=\mathcal{L}(q,w_\mu)\eeq Now, suppose
that $\rho(q,x)$ and $j_\mu(q,x)$ describe an actual evolution of our system in the space-time.
We consider the family of subsets $\Omega_+=\mathrm{U}\times\mathrm{Z}$, where $\mathrm{U}$ and
$\mathrm{Z}$ are compact domains of $\mathbb{R}^n$ and $\mathrm{M}$, respectively, such that
$\rho(q,x)>0$ for $q,x \in \Omega_+$. In each $\Omega_+$, we make the substitution (analogous
to \rif{11}) \beq w_\mu \, \rightarrow \, \frac{j_\mu (q,x)}{\rho(q,x)} \nonumber \eeq in the
given Lagrangian \rif{88} and then we consider the action functional \barr \label{eq89}
I_{\Omega_+}[j_\mu,\rho,\lambda^\sigma]&=&\int_{\Omega_+}dxdq\biggr[\mathcal{L}(q,\frac{j_\mu
(q,x)}{\rho(q,x)})\,\rho(q,x)+ \nonumber \\ &+&\lambda^\nu \bigr(\partial_\nu \rho(q,x) + \dpr
k j_\nu^k(q,x)\bigr) \biggr]\,, \earr where, in order to take into account of the equations
\rif{87}, we have introduced four Lagrange multiplier fields $\lambda^\mu(q,x)$, defined in
$\Omega_+$. Now we demand that the actual $\rho(q,x)$ and $j_\mu(q,x)$ are such that
$I_{\Omega_+}[j_\mu,\rho,\lambda^\sigma]$ is stationary under smooth independent variations
$\delta\rho(q,x)$ of $\rho(q,x)$ and $\delta j_\mu(q,x)$ of $j_\mu(q,x)$, with
$\delta\rho(q,x)=0$ for $q,x \in \partial U\times\partial Z$ and $\delta j_\mu(q,x)=0$ for $q
\in \partial U$. Furthermore the equations \rif{87} must hold.

By considering the variation of $j_\mu(q,x)$ we obtain from \rif{89} \barr \label{eq90} \dpr k
\lambda^\nu(q,x) &=& \de{\mathcal{L}}{w_\nu^k}\biggr(q,\frac{j_\mu (q,x)}{\rho(q,x)}\biggr)
\nonumber \\ &&\quad (k=1,2,..,n\,;\,\nu = 0,1,2,3) \earr

The variation of $\rho$ gives \barr \label{eq91} \dpr \nu \lambda^\nu(q,x) &+& \frac{j_\mu
(q,x)}{\rho(q,x)}\de{\mathcal{L}}{w_\nu^k} \biggr(q,\frac{j_\mu (q,x)}{\rho(q,x)}\biggr)-
\nonumber \\ &-& \mathcal{L}\biggr(q,\frac{j_\mu (q,x)}{\rho(q,x)}\biggr) = 0 \earr
 Now let us assume that \beq \label{eq92} det
 \biggr[\frac{\partial^2 \mathcal{L}}{\partial w_\nu^i \partial
 w_\sigma^j} (q,w_\mu) \biggr] \neq 0 \eeq

 Then we can solve \rif{90} with respect to ${j_\mu}\bigr/\rho$. Through this solution the
equations \rif{87} and \rif{91} can be expressed in an appropriate and useful form, by making
use again of the Legendre transformation.

 Let us introduce a generalized "Hamiltonian" function $\mathcal{H}$
\beq \label{eq93} \mathcal{H} = \mathcal{H}(q,\pi^\mu) = w_\nu^k \pi_k^\nu -
\mathcal{L}(q,w_\mu) \quad \bigr(\pi^\mu = (\pi_1^\mu,..,\pi_n^\mu)\bigr)\,,\eeq where
$w_\nu^k(q,\pi^\mu)$ is the solution of the system of equations \beq \label{eq94} \pi_i^\nu =
\de{\mathcal{L}}{w_\nu^i}(q,w_\mu) \eeq Then we have \beq \label{eq95} \frac{j_\nu^k}{\rho} =
 \de{\mathcal{H}}{\pi^\nu_k}(q,\partial \lambda^\mu) \quad (\partial \lambda^\mu = (\partial_1 \lambda^\mu,..,\partial_n \lambda^\mu) \eeq
 So we obtain the system of equations \beq \label{eq96} \dpr \nu
 \lambda^\nu(q,x) + \mathcal{H}(q,\partial \lambda^\sigma (q,x)) =
 0\eeq \barr \label{eq97} \dpr \mu \rho(q,x) &+& \dpr k \biggr(\rho
 \de{\mathcal{H}}{\pi^\mu_k}(q,\partial \lambda^\sigma (q,x))\biggr) =
 0\nonumber \\ &&
 \qquad (q,x \in \Omega_+) \earr
 Due to the absence of an explicit restriction connected with the
set of points in $\mathbb{R}^n \times M$ where $\rho(q,x)=0$, we can reconsider \rif{96} and
\rif{97} as equations in $\mathbb{R}^n \times M$, free from constraints. They are the version,
in the framework of the previous probabilistic scheme for classical field theory, of the
equations \rif{21} and \rif{22} for a m.s.f. Through \rif{96} we have obtained a generalization
to field theory of the Hamilton-Jacobi equation. However this generalization is again a partial
differential equation, and not a functional equation, as happens in the standard canonical
field theory. Furthermore the space coordinates are treated on the same
 footing of the time. These four parameters together take the place
 that the time alone had in previous sections. The system
 \rif{96},\rif{97} satisfies manifestly relativistic invariance,
 if $\mathcal{L}$, as we will always assume, is a Lorentz scalar.
 The generalized "Hamiltonian" $\mathcal{H}$, given by \rif{93},
 is not an energy density, but a Lorentz scalar. Another feature
 of our approach to field theory, is the occurrence of a system of
 four equations for $\rho$, in place of the only one equation
 \rif{22} for a m.s.f. As a consequence \rif{96} is now not
 decoupled from \rif{97}. The functions $\lambda^\mu(q,x)$,
 besides to satisfy \rif{96}, have to guarantee the consistency of
 the system \rif{97}.

 The Euler-Lagrange equations of the standard deterministic classical field theory, can be considered
as a particular consequence of the system \rif{96},\rif{97}.

 Let $q(x)=(q^1(x),..\,,q^n(x)))$ be a classical field. It can be
 easily verified that \rif{97} admits as a particular solution
 the probability density \beq \label{eq98} \rho(q,x) =
 \delta(q-q(x)) \eeq if, and only if, the field $q(x)$ is such
 that \beq \label{eq99} \dpr \sigma q^k(x) =
 \de{\mathcal{H}}{\pi_k^\sigma}\biggr(q(x),\partial
 \lambda^\mu(q(x),x)\biggr) . \eeq On the other hand, by deriving \rif{96} with respect to $q^i$, we see that
 $q(x)$ must satisfy also the equation \beq \label{eq100} \dpr \nu
 \dpr i \lambda^\nu\bigr(q(x),x\bigr) = -\de{\mathcal{H}}{q^i}\biggr(q(x),\partial
 \lambda^\mu(q(x),x)\biggr) \eeq
 By taking into account of the Legendre transformation \rif{93},
 we deduce from \rif{99} \beq \label{eq101} \dpr i
 \lambda^\mu(q(x),x)= \de{\mathcal{L}}{w_\mu^i}\bigr(q(x),\dpr \nu
 q(x) \bigr) \equiv \pi_i^\mu (x)  \eeq

 Then the equation \rif{100} gives \beq \dpr \nu \biggr(\de{\mathcal{L}}{w_\nu^i}\bigr( q(x),\dpr \mu
 q(x)\bigr)\biggr) - \de{\mathcal{L}}{q^i}\bigr( q(x),\dpr \mu
 q(x)\bigr) = 0 \,,\nonumber \eeq
that is the Euler-Lagrange equations \rif{3} for the field $q(x)$.

 A function $q(x)$ which is a solution of \rif{99}, with
 $\lambda^\mu (q,x)$ solution of \rif{96}, is said to be
 embedded in a geodesic field
 \cite{bib_2,bib_6,bib_25}. As we have seen, such a
function is an extremal, i. e. a solution of \rif{3}. It can be shown that any extremal $q(x)$
can be embedded in a geodesic field, in some neighborhood of the surface $q=q(x)$
\cite{bib_1,bib_2,bib_6,bib_7,bib_25}.

 We can take directly as unknown the fields $\dpr i
\lambda^\mu(q(x),x) = \pi_i^\mu(x)$. Then we can consider directly the system \beq
\label{eq102} \dpr \mu q^i(x) = \de{\mathcal{H}}{\pi_i^\mu}\biggr(q(x),\pi^\nu(x)\biggr)\eeq
\beq \label{eq103} \dpr \mu \pi_i^\mu(x) =
-\de{\mathcal{H}}{q^i}\biggr(q(x),\pi^\nu(x)\biggr)\,,\eeq which is the relativistic covariant
version, in classical field theory, of the canonical Hamilton's equations.

 As a matter of fact, the equations \rif{96},\rif{99},
\rif{102},\rif{103}, which we have deduced through our probabilistic approach, have already
been obtained on the basis of different considerations, in the deterministic De Donder-Weyl
theory developed for multi-dimensional variational problems
\cite{bib_1,bib_2,bib_6,bib_7,bib_8,bib_14,bib_25}. In the framework of the deterministic
classical field theory, they have been also reobtained elsewhere \cite{bib_26,bib_27}.

 We note that, in the De Donder-Weyl theory for classical fields,the relevant aspect of manifest relativistic covariance compels
to have four "momentum density" $\pi_i^\mu(x)$ conjugate to each classical field $q^i(x)$. This
makes problematic the standard Poisson brackets formalism and then the transition to a quantum
theory. However this redundance of conjugate variables $\pi_i^\mu(x)$can be reduced through
proper constraints. As a matter of fact these constraints are contained in the equations
\rif{102}.

 We can consider a family of space-like surfaces which covers
 simply the manifold $M$. Such a family can be related to a
 description of the dynamical evolution of our system. However, in
 the following, for simplicity, we will limit ourselves to
 surfaces having $x_0$ constant.

 Now the three equations in \rif{102} \beq \label{eq104} \dpr m q^i(x) =
 \de{\mathcal{H}}{\pi_i^m}\biggr(q(x),\pi^0(x),\pi^n(x)\biggr) \qquad
 (m = 1,2,3)\eeq can be interpreted as constraints for the fields
 $\pi^n(x) \quad (n=1,2,3)$.

 By solving them, we will obtain \beq \label{eq105} \pi^m_i(x)=
F_i^m \biggr(q(x), \pi^0(x),\dpr n q(x)\biggr). \eeq Then we can introduce the function \barr
\label{eq106} && \mathcal{H}_c\biggr(q(x), \pi^0(x),\dpr m q(x)\biggr) = \nonumber \\ &&\quad=
\mathcal{H}\biggr(q(x),
\pi^0(x),F^m\bigr(q(x), \pi^0(x),\dpr n q(x)\bigr)\biggr) - \nonumber \\
&&\quad - \bigr(\dpr m q^i(x) \bigr) F_i^m \biggr(q(x), \pi^0(x),\dpr n q(x)\biggr)\earr
 We have \beq \label{eq107} F_i^m\bigr(q(x),\pi^0(x),\dpr n
 q(x)\bigr) =- \de{\mathcal{H}_c}{(\dpr m q^i(x))} \eeq
 The system \rif{102},\rif{103} is then reduced to \beq
 \label{eq108} \dpr 0 q^i(x) = \de{\mathcal{H}_c}{\pi^0_i} \eeq
 \beq \label{eq109} \dpr 0 \pi^0_i(x) = -
 \biggr(\de{\mathcal{H}_c}{q_i}-\dpr m \de{\mathcal{H}_c}{(\dpr m
q^i(x))}\biggr), \eeq that is the system of equations of the standard Hamiltonian formalism
\cite{bib_4}, related to the description of a classical field as a m.s.i. Really, it follows
from \rif{93} that $\mathcal{H}_c$ is the standard Hamiltonian density.

 Then we see that, if we carry out the elimination of the
 conjugate fields $\pi_i^n(x)$, the resulting theory (which is the
 standard one), meets with the difficulties mentioned at the beginning
 of this section.
 On the other hand, if we keep the complete manifest covariant
 structure of the De Donder-Weyl theory, then this theory, as we
 have shown, can be embedded in a simple and well defined
 probabilistic scheme, related to a relativistic invariant
 generalized partial differential equation of Hamilton-Jacobi
 type. In such a scheme it is possible to contemplate quantum
 extensions, as in the previous section, which are disengaged from
 the usual s.c.q.r. This will be done in the next section.

 Due to \rif{89} and \rif{95}, the basic equations of the previous
 scheme, that is \rif{96} and \rif{97}, can be derived directly
 from the standard variational principle applied to the action
 integral \beq \label{eq110} A_\Omega[\rho,\lambda^\mu] =
 \int_\Omega dx dq \biggr[- \lambda^\nu(q,x)\dpr \nu
 \rho(q,x)+\mathcal{H}(q,\partial \lambda^\mu)\rho(q,x)\biggr]\,,
 \eeq
 where $\Omega$ is now a generic domain in of $\mathbb{R}^n\times
 M$. The variations of $\lambda^\mu$ must vanish on the boundary
 of $\Omega$. From the point of view of standard classical field
 theory, we see then that only $\lambda^0(q,x)$ has a conjugate
 variable, that is $\rho(q,x)$, while for $\lambda^m(q,x) \quad
 (m=1,2,3)$such conjugate variables are absent. Therefore three of
 the equations \rif{97}, that is \beq \label{eq111} \dpr m
 \rho(q,x) + \dpr k \biggr(\rho
 \de{\mathcal{H}_l}{\pi^m_k}\bigr(q,\partial
 \lambda^\sigma(q,x)\bigr)\biggr) = 0 \qquad (m=1,2,3) \eeq
 have to be considered as constraints of the dynamical evolution,
 in analogy with what happens for the system \rif{102}. We note that the constraints which
 appear in the previous approach are intrinsically related to the property that the generalized
 hamiltonian is a scalar, and not an energy density. As a matter of fact, due to \rif{92}, we
 are not treating singular Lagrangians. It seems that these constraints have a role different
 from that given by the Dirac approach.

We conclude this section with the following remarks. First of all, the previous probabilistic
approach to classical field theory requires, in general, some caution. This is related to the
condition \rif{92}, which makes possible a transition from the Lagrangian formalism to a
covariant Hamiltonian formalism. In the case of a classical Dirac field, it is well known the
Lagrangian density is linear in the derivatives of the field components so that the condition
\rif{92} cannot be satisfied. Another problem, of technical character, appears if we have
constraints between the components of classical fields, as happens in the case of vector
fields. The role and meaning of these cases in our probabilistic approach will be investigated
elsewhere. In the following we will limit our considerations to scalar fields having the
property to satisfy the condition \rif{92}.

Since we assumed that our classical Lagrangian density \rif{1} does not depend explicitly on
$x$, we can consider the invariance of our probabilistic theory under space-time translations,
that is under the transformation \beq \label{eq112} y=(x,q)\,\rightarrow\,y' =
(x+\varepsilon,q) \quad
\varepsilon=(\varepsilon^0,\varepsilon^1,\varepsilon^2,\varepsilon^3)\eeq
 Then, from \rif{112} and a straight-forward application of the
Noether's theorem to the action \rif{110}, we deduce the local conservation law \beq
\label{eq113} \dpr \mu
 \bigr(T^\mu_\nu(q,x)\rho(q,x)\bigr) +\dpr i Q^i_\nu(q,x) = 0\,,\eeq
 with \beq \label{eq114} T^\sigma_\nu(q,x) = \dpr k
 \lambda^\sigma(q,x) \de{\mathcal{H}}{\pi^\nu_k}(q,\partial
 \lambda^\mu) + g^\sigma_\nu \biggr(\mathcal{H}-\dpr k
 \lambda^\alpha \de{\mathcal{H}}{\pi^\alpha_k}\biggr)\,.\eeq
$Q^i_\nu$ is a function of $q$,$\rho(q,x)$,$\partial \lambda^\mu(q,x)$.

 So, our approach contemplates a random energy-momentum tensor
 $T_\nu^\mu(q,x)$ given by \rif{114}. If we consider the mean of
 $T_\nu^\mu(q,x)$at the point $x$ \beq \label{eq115}
 \overline{T}_\nu^\mu(q,x)= \int_{\mathbb{R}^n}
 T_\nu^\mu(q,x)\rho(q,x)dq \eeq we deduce from \rif{113} \beq
 \label{eq116} \dpr \mu \overline{T}^\mu(x) = 0\eeq
 In the case of the solution \rif{98}, we obtain from \rif{114}
 and \rif{115} \barr \label{eq117} \overline{T}_\nu^\sigma(x) &=&
 {T_c}_\nu^\sigma(x) = \nonumber \\ &=&
 \de{\mathcal{L}}{w^i_\sigma}\bigr(q(x),\dpr \mu q(x)\bigr)\dpr
 \nu q^i(x)-g^\sigma_\nu \mathcal{L}\bigr(q(x),\dpr \mu q(x)
 \bigr)\,, \nonumber \\ \earr
i. e. the standard result of the deterministic classical field theory.

\section{\label{sec:level5}The quantum extension for scalar fields}

We develop a quantum theory for scalar fields, starting from the probabilistic scheme of the
previous section and following the approach of the sect.\ref{sec:level3}. We will limit our
considerations to a set of $n$ real fields described by a "natural" Lagrangian function \beq
\label{eq118} \mathcal{L}(q,w_\mu) = \frac 1 2 \eta_{ij}(q)w_\nu^i w^{\nu j}-V(q)\eeq where
$\eta_{ij}(q)$ is positive definite $\forall q \quad (\eta_{ij}(q)=\eta_{ji}(q)$, the inverse
matrix is denoted $\eta^{ij}(q)$).The Hamiltonian function \rif{93} in this case becomes \beq
\label{eq119} \mathcal{H}(q,\pi^\mu) = \frac 1 2 \eta^{ij}(q)\pi_{\nu i} \pi_j^\nu + V(q) \eeq
As a first step of an extension of the scheme of the previous section, we take into account, in
the balance equation \rif{87}, of an additional contribution to $\dpr \mu \rho(q,x)$. This is
given by current densities $i^k_\mu(q,\rho(q,x),\dpr i \rho(q,x),...)$ having a local
dependence on $\rho$ and on its derivatives with respect to $q^i$. So the equation \rif{87} is
extended to \beq \label{eq120} \dpr \mu \rho(q,x) + \dpr k j_\mu^k(q,x) + \dpr k
i_\mu^k\bigr(q,\rho(q,x),\dpr j \rho(q,x),..\bigr)=0 \eeq Through $i_\mu^k$ we introduce
dynamical mechanisms of diffusive type. These mechanisms are internal so that, due to the
homogeneity of space-time, they are assumed to have no explicit dependence on $x$. However, due
to the structure of $i_\mu^k$, we are faced also with the problem to assure that, in the theory
which we are developing, the isotropy of space-time is not destroyed. To this end, we note
that, as in the case of a m.s.f., the transition from \rif{87} to \rif{120} makes relevant the
choice of a Lagrangian density which, at the classical level, is equivalent to
$\mathcal{L}\bigr(q,{j_\mu(q,x)}\bigr/{\rho(q,x)}\bigr)$. As a matter of fact, when \rif{87} is
satisfied, the Lagrangian density \beq \label{eq121} \widetilde{\mathcal{L}} =
\mathcal{L}\biggr(q,\frac{j_\mu}{\rho}\biggr) + D(\rho)a^\nu \frac{j^k_\nu(q,x)}{\rho(q,x)}\dpr
k \rho(q,x)\, , \eeq where $D(\rho)$ is a given function of $\rho$ and $a^\nu$ is a given
constant four vector, is equivalent to $\mathcal{L}$.

Then, as a basis for a quantum theory of our system, we consider the action functional \barr
\label{eq122}
\widetilde{I}_{\Omega_+}[j_\mu,\rho,\lambda^\sigma]&=&\int_{\Omega_+}dxdq\biggr\{\biggr[\mathcal{L}(q,\frac{j_\mu
}{\rho})+ D(\rho)a^\nu \frac{j^k_\nu}{\rho}\dpr k \rho \biggr]\,\rho \nonumber \\
&+& \lambda^\nu \bigr(\partial_\nu \rho + \dpr k j_\nu^k + \dpr
 k i_\nu^k(q,\rho,\dpr l \rho, .. )\bigr) \biggr\}\nonumber \\ \earr

 The equations \rif{121} and \rif{122} are the generalization to
 our fields of the equations \rif{26} and \rif{31} for mechanical
 systems.

 In the case of non relativistic mechanical systems we were faced
with the breaking of time-reversal invariance, which is related to the choice of an arrow of
the time axis. In the case of fields, the introduction of a constant four vector $a^\nu$, that
is a privileged direction in the Minkowski space, breaks the isotropy of the space-time. At the
classical level we have no problem, since the additional term in the equation \rif{121} is
ineffective. This is not the case, in general, if we start from the action functional
\rif{122}. However the isotropy of the space-time can be still preserved by choosing
appropriately the current $i_\mu^k$.
 As a matter of fact, the request for the space-time isotropy allows
to determine completely $i_\mu^k$, starting from the knowledge of $\widetilde{\mathcal{L}}$. In
fact, on the basis of this request, we deduce from \rif{118},\rif{121} and \rif{122} that \beq
 \label{eq123} i_\nu^k(q,\rho,\dpr l \rho, .. )=
a_\nu\eta^{kj}(q)\rho D(\rho)\dpr j \rho \,,\eeq which is a generalization of \rif{41}.

Then we deduce from \rif{118} and \rif{122} the Lorentz invariant system of equations \barr
\label{eq124} &&\dpr \nu \lambda^\nu + \frac 1 2 \eta^{ij}\dpr i \lambda_\nu \dpr j \lambda^\nu
+ V(q) + \nonumber \\&&+ a_\nu a^\nu\bigr[\frac 1 2 \eta^{ij}\bigr(\rho D^2(\rho)\bigr)' \dpr i
\rho \dpr j \rho - \dpr i\bigr(\eta^{ij}\rho D^2(\rho)\dpr j \rho \bigr)\bigr] = 0 \nonumber \\
\earr \beq \label{eq125} \dpr \mu \rho + \dpr k \bigr(\rho \eta^{kj}\dpr j \lambda_\mu\bigr) =
0 \quad \bigr((\mu=0,1,2,3)\,,\,(q,x \in \Omega_+)\bigr), \eeq which appears as a
straight-forward generalization to fields of the system \rif{44},\rif{45}, obtained in the
quantum theory of a m.s.f.

The system \rif{124},\rif{125} can be derived directly from the action integral \barr
\label{eq126} A_{\Omega_+}[\rho,\lambda^\mu]&=&\int_{\Omega_+}dxdq\biggr[ -\lambda^\nu
(q,x)\partial_\nu \rho(q,x) + \mathcal{H}(q,\partial \lambda^\mu)\rho  \nonumber \\ &+&
\frac{a_\nu a^\nu}{2} \rho D^2(\rho)\eta^{kl}\dpr k \rho \dpr l \rho \biggr] \,,\earr where
$\mathcal{H}$ is given by \rif{119} (the variations of $\lambda^\mu$ must vanish on the
boundary of $\Omega_+$).

The last term in the equation \rif{126} has the same structure which we have found in the case
of a m.s.f. So we can speak for fields of an extended classical domain or of a quantum domain,
according to the behavior of $\rho D^2(\rho)$. We will fix our attention on the quantum domain.
This will be characterized, as in the case of mechanical systems $\bigr(eq.\rif{51}, g(\rho) =
0\bigr)\,,$ by requiring \beq \label{eq127} a_\nu a^\nu \rho D^2(\rho) = \frac{(f/2)^2}{\rho}
\qquad (f>0)\,, \eeq where $f$ is constant. We have assumed in \rif{127} that $a_\nu a^\nu >0$,
i. e. the currents $i_\nu^k$ are time-like, as can be expected on physical grounds.

In the quantum domain \rif{124} becomes \barr \label{eq128}  && \dpr \nu \lambda^\nu + \frac 1
2 \eta^{ij}\dpr i \lambda_\nu \dpr j \lambda^\nu + V(q) + \nonumber
\\&&+ \frac {f^2}{4}\biggr(\frac 1 2 \eta^{ij} \frac{\dpr i \rho\dpr j
 \rho}{\rho^2}-\frac{1}{\rho}\dpr i \bigr(\eta^{ij}\dpr i
 \rho\bigr)\biggr) = \nonumber \\ &&= \dpr \nu \lambda^\nu + \frac 1 2 \eta^{ij}\dpr i
\lambda_\nu \dpr j \lambda^\nu + V(q) - \frac {f^2}{2} \frac{\dpr i ( \eta^{ij}\dpr j
\rho^{1/2})}{\rho^{1/2}}= \nonumber \\ && = 0\,,\qquad (q,x)\in \Omega_+ \earr The system
\rif{125},\rif{128} has a Lorentz invariant local hydrodynamic form, which generalizes the
Madelung hydrodynamic equations related to non relativistic quantum mechanics. The system is
also invariant under the transformations involving space-time inversion \barr \label{eq129}
x^\mu & \rightarrow & x'^\mu = -x^\mu \nonumber \\
q^i & \rightarrow & q'^i = q^i \nonumber \\
\rho(q,x^\mu) & \rightarrow & \rho'(q,x'^\mu)=\rho(q,x^\mu) \nonumber \\
\lambda^\mu(q,x^\nu) & \rightarrow & \lambda'^\mu(q,x'^\nu) =
-\lambda^\mu(q,x^\nu)  \earr

We note that, while in the case of mechanical systems the constant $a$ of eq.\rif{51},
analogous to $f$, has the dimension of an action (as a matter of fact $a = \hbar$, according to
\rif{68}), in the case of our fields the constant $f$ has the dimension of a (spatial) density
of an action, as a consequence of our previous equations \rif{126},\rif{127}. This means that,
by considering $\hbar$, $f$ can be related to a fundamental quantity having the dimension of a
spatial length.

We will see some consequences of the system \rif{125},\rif{128}, by assuming that, as in some
usual models of field theory (Klein-Gordon, $\lambda q^4,...$), \beq \label{eq130} V(q) \geq 0
\, , \, V(q)\rightarrow +\infty \; for |q|\rightarrow +\infty \eeq

Furthermore we will consider the random energy-momentum tensor ${T_Q}^\sigma_\nu(q,x)$ in the
quantum case. This can be obtained by adding to the classical expression \rif{114} the "quantum
potential" which appears in the equation \rif{128}. Taking into account of \rif{119}, we have
\barr \label{eq131} {T_Q}^\sigma_\nu(q,x)&=& \eta^{ij}(q)\dpr i \lambda^\sigma\dpr j
\lambda_\nu + g^\sigma_\nu\biggr( -\frac {f^2}{2} \frac{\dpr i ( \eta^{ij}\dpr j
\rho^{1/2})}{\rho^{1/2}} + \nonumber \\ &+& V(q)- \frac 1 2 \eta^{ij}\dpr i \lambda_\mu \dpr j
\lambda^\mu\biggr),\earr for $ (q,x)\in \Omega_+$.

In particular, the random energy density $\varepsilon^Q(q,x)$ and the random momentum densities
$P^Q_m(q,x) \quad (m=1,2,3)$, are given by \barr \label{eq132} \varepsilon^Q(q,x) =
{T_Q}^0_0(q,x) &=& \frac 1 2 \eta^{ij}\dpr i \lambda^0 \dpr j \lambda^0 + \frac 1 2
\eta^{ij}\dpr i \lambda^m \dpr j \lambda^m +\nonumber \\ &+& V(q) - \frac {f^2}{2} \frac{\dpr i
( \eta^{ij}\dpr j \rho^{1/2})}{\rho^{1/2}} \earr \beq \label{eq133} P^Q_m(q,x) = {T_Q}^0_m(q,x)
= \eta^{ik}\dpr k \lambda^0 \dpr i \lambda_m \qquad (m=1,2,3) \eeq

The simplest solutions of the system \rif{125},\rif{128} can be obtained by considering the
case \beq \label{eq134} \dpr i \lambda_\mu(q,x) = 0 \qquad (i=1,2,..,n\,;\,\mu=0,1,2,3)\eeq

In such a case, \rif{125} is solved automatically by $\rho(q,x)$ independent of $x$, that is
$\rho(q,x) = \widetilde{\rho}(q)$. Furthermore \rif{128} (with $\tilde{\rho}$) requires
necessarily that also $\dpr \nu \lambda^\nu$ is independent of $x$, that is, due to \rif{134},
\beq \label{eq135} \dpr \nu \lambda^\nu (q,x) = -w \,,\eeq where $w$ is a constant. Then
\rif{128} becomes \beq \label{eq136} - \frac {f^2}{2} \frac{\dpr i ( \eta^{ij}\dpr j
\widetilde{\rho}^{1/2})}{\widetilde{\rho}^{1/2}}+V(q) = w \eeq for $q$ such that
$\widetilde{\rho}(q)>0$.

This equation can be translated in a global form, by setting \beq \label{eq137}
\widetilde{\rho}(q) = \psi^2(q)\,, \eeq so that we are led to replace \rif{136} with \beq
\label{eq138} \biggr(- \frac {f^2}{2} \dpr i ( \eta^{ij}\dpr j
\psi(q))\biggr)+V(q)\psi(q)=w\psi(q)\,, \eeq that is a stationary Schr\"{o}dinger equation with
the constant $f$, instead of $\hbar$

Let us denote by $w_r$ the eigenvalues of the "Hamiltonian" operator \beq \hat{H} = - \frac
{f^2}{2}\dpr i ( \eta^{ij}\dpr j \cdot)+V(q). \nonumber \eeq We order them according to \beq 0
\leq w_0 < w_1 < ... \nonumber \eeq

We conclude that our system \rif{125},\rif{128} admits as solutions invariant states, which are
associated to the eigenvalues $w_r$ of $\hat{H}$. In an invariant state, corresponding to a
particular eigenvalue $w_s$, the energy momentum tensor is well defined, constant and given by
\beq \label{eq139} {T_Q}^\sigma_\nu = g^\sigma_\nu w_s \eeq

Due to \rif{139}, we see that it is a consequence of our equations the existence of several
non-degenerate vacuum states, each having a constant finite energy density. We will call the
state corresponding to $w_0$ the fundamental vacuum, while the other states associated to $w_r
\quad (r>0)$ may be called virtual vacua. The energy gap between the fundamental vacuum and a
virtual vacuum is infinite. We note that, in our approach, there is no substantial distinction
between the case of interacting fields and the non interacting one. The fundamental vacuum
exists also for interacting fields. In the traditional approach, in the case of interacting
fields, the existence of the vacuum state must be, in some way, postulated. In the case of free
fields, the traditional approach gives an infinite zero point energy, but this is infinite also
in any finite spatial volume, while our $w_0$ is finite.

The previous aspects of our approach may be relevant in connection with the dark energy problem
\cite{bib_28}. We note also that, as a consequence of \rif{130}, the eigenfunctions $\psi_r(q)$
associated to the invariant states are rapidly decreasing for $|q|\rightarrow +\infty$. Then we
have that the fluctuations \beq \label{eq140}
\overline{(q_i-\overline{q}_i)(q_j-\overline{q}_j)} = \int dq
(q_i-\overline{q}_i)(q_j-\overline{q}_j)\psi^2_0(q) \eeq ($\bar{q}_i = \int dq q_i
\psi_0^2(q)$, the $\psi_r(q)$ are supposed normalized) are finite. In the traditional approach,
if we consider, for example, the real Klein-Gordon field, the analogous quantity $\bra 0
\hat{q}^2(x)\ket 0$ ($\hat{q}$ the operator valued Heisenberg field) is infinite.

To simplify our further discussion, in the following we will limit ourselves to the case of
only one scalar field, assuming also that $\eta_{ij}(q)=\eta(q)=\eta$ (positive constant).

Now, as next steps, we will examine space-independent solutions and time-independent solutions.
The first type corresponds to solutions such that \beq \label{eq141} \de{}{q} \lambda^m(q,x) =
0 \qquad (m=1,2,3) \eeq but $\partial\lambda^0\bigr/\partial q \neq 0$, in general.

With $\rho(q,x)=\tilde\rho(q,x^0)$, as a consequence of \rif{141}, the system
\rif{125},\rif{128} becomes \beq \label{eq142} \de{}{x^0}\tilde{\rho}(q,x^0)+\frac 1 \eta
\de{}{q}\biggr(\tilde{\rho}\de{}{q}\lambda_0(q,x)\biggr) = 0\eeq \barr \label{eq143}
&&\de{}{x^0}\lambda_0(q,x)+\de{}{x^m}\lambda^m(q,x)+\frac{1}{2\eta}\bigr(\de{}{q}\lambda_0(q,x)\bigr)^2
+  \nonumber \\
&& +V(q) -\frac{f^2}{2\eta}\frac{1}{\tilde{\rho}^{1/2}} \frac{\partial^2
\tilde{\rho}^{1/2}}{\partial q^2} = 0\earr

From \rif{142} we deduce $\lambda^0(q,x)=\lambda(q,x^0)$, while \rif{141} and \rif{143} require
that $\de{}{x^m}\lambda^m(q,x) = f(x^0)$. Writing  \beq f(x_0) = \de{}{x^0} \int^{x^0} f(x'^0
)dx'^0 \equiv \de{}{x^0}\Lambda(x^0) \; , \nonumber \eeq we can absorb $\Lambda(x^0)$ within
$\lambda(q,x^0)$, by considering $\widetilde{\lambda}(q,x^0)= \lambda(q,x^0)+\Lambda(x^0)$.
Then we arrive at the equations \begin{equation*}\tag{\ref{eq142}'} \label{eq142'}
\de{}{x^0}\tilde{\rho}(q,x^0)+\frac 1 \eta
\de{}{q}\biggr(\tilde{\rho}\de{}{q}\tilde{\lambda}(q,x^0)\biggr) = 0 \end{equation*}
\begin{equation*}\tag{\ref{eq143}'}
\label{eq143'} \de{}{x^0}\tilde{\lambda}(q,x^0)+\frac{1}{2\eta}
\biggr(\de{}{q}\tilde{\lambda}(q,x^0)\biggr)^2 + V(q)
-\frac{f^2}{2\eta}\frac{1}{\tilde{\rho}^{1/2}}\frac{\partial^2 \tilde{\rho}^{1/2}}{\partial
q^2} = 0
\end{equation*}

According to sect.\ref{sec:level3}, this system is the local form of the global linear
Schr\"{o}dinger equation \beq \label{eq144} \emph{i}f \de{}{x^0} \widetilde{\psi}(q,x^0)= -
\frac{f^2}{2\eta}\de{}{q^2}\widetilde{\psi}(q,x^0)+V(q)\widetilde{\psi}(q,x^0) \,,\eeq where
locally \beq
\widetilde{\psi}(q,x^0)=\tilde{\rho}^{1/2}(q,x^0)exp\biggr(\frac{\emph{i}}{f}\tilde{\lambda}(q,x^0)\biggr)\;
. \nonumber \eeq

The general solution of \rif{144} is a linear superposition of the vacuum states. Due to
\rif{133} and \rif{141}, we have for this solution, $P^Q_m(q,x)=0 \quad (m=1,2,3)$, while the
random energy density is, in general, time-dependent. On the other hand, its expectation value
is a constant. In fact, due to \rif{143'}, \rif{133} gives \beq \label{eq145}
\varepsilon^Q(q,x^0)=-\de{}{x^0}\tilde{\lambda}(q,x^0)=\emph{i}\frac f 2
\frac{1}{\tilde{\psi}\tilde{\psi}^\ast}\bigr(\tilde{\psi}^\ast\de{}{x^0}\tilde{\psi}-\tilde{\psi}\de{}{x^0}\tilde{\psi}^\ast\bigr)\,,
\eeq so that \beq \label{eq146} \int dq \varepsilon^Q(q,x^0)\tilde{\rho}(q,x^0) = \bar{w}=\int
dq \tilde{\psi}^\ast(q,x^0)\emph{i}f\de{}{x^0}\tilde{\psi}^\ast(q,x^0) \eeq

Now we discuss the very interesting case of time-independent solutions, by considering the
situation \beq \label{eq147} \de{\lambda^0}{q}(q,x) = 0\,,\eeq while, in general,
$\de{}{q}\lambda^m(q,x)\neq0\,.$ Also in this case $P^Q_m(q,x)=0$, but now there appear
important features of the random energy density.

As a consequence of \rif{147} we have
$\rho(q,x)=\hat\rho(q,\vec{x})$,$\lambda^m(q,x)=\hat\lambda^m(q,\vec{x})$,
$\bigr(\vec{x}=(x^1,x^2,x^3),m=1,2,3\bigr)$ while, taking into account of \rif{128}, we can
write $\de{}{x^0}\lambda^0(q,x)=g(\vec x)$. By introducing $h^m(\vec x)$, such that
$\de{}{x^m}h^m(\vec x)=g(\vec x)$, we can absorb $h^m(\vec x)$ within $\hat\lambda^m(q,\vec
x)$, by considering $\hat\lambda^m(q,\vec x)+h^m(\vec x)$, which we recall
$\hat\lambda^m(q,\vec x)$. Our system \rif{125},\rif{128} becomes \beq \label{eq148}
\de{}{x^m}\hat\rho(q,\vec x)+\frac 1 \eta \de{}{q}\biggr(\hat\rho(q,\vec
x)\de{}{q}\hat\lambda_m(q,\vec x)\biggr)=0 \quad (m=1,2,3) \eeq \barr \label{eq149}
&&\de{}{x^m}\hat\lambda^m(q,\vec x)-\frac{1}{2\eta}
\sum_{m=1}^3\biggr(\de{}{q}\hat\lambda_m(q,\vec x)\biggr)^2+\nonumber
\\ &&+ V(q)-\frac{f^2}{2\eta \hat\rho^{1/2}}\frac{\partial^2}{\partial
q^2}\hat\rho^{1/2}(q,\hat x)=0\earr

In order to see some basic aspect of this system, we fix the
attention on its possible spherical symmetric solutions. We start
from the ansatz \beq \label{eq150} \hat\lambda^m(q,\vec
x)=-\hat\lambda^m(q,r)\frac{x^m}{r} \qquad (r=|\vec
x|,m=1,2,3)\eeq

 We infer then from \rif{148} $\hat\rho(q,\vec x)=\hat\rho(q,r)$.
 The system \rif{148},\rif{149} becomes \begin{equation*}\tag{\ref{eq148}'}
\label{eq148'}\de{}{r}\hat\rho(q,r)+\frac 1 \eta
\de{}{q}\biggr(\hat\rho(q,r)\de{\hat\lambda(q,r)}{q}\biggr)=0
 \end{equation*} \begin{align*}
\label{eq149'} -\de{}{r}\hat\lambda(q,r)-\frac 2 r \hat\lambda(q,r)-\frac{1}{2\eta}
\biggr(\de{}{q}\hat\lambda(q,r)\biggr)^2+V(q)-\\-\frac{f^2}{2\eta\hat\rho^{1/2}(q,r)}\frac{\partial^2}{\partial
q^2}\hat\rho^{1/2}(q,r)=0 \tag{\ref{eq149}'}
 \end{align*}
This cannot be translated in a Schr\"{o}dinger type equation, due to the minus sign in the
third term of \rif{149'}. However it is useful to introduce a transformation involving two
real, "conjugate", functions, as is done in diffusion theory \cite{bib_21}. We consider two
real functions, $\phi(q,r)$ and $\tilde\phi(q,r)$, such that \beq \label{eq151} \rho(q,r) =
\tilde\phi(q,r) \phi(q,r)\,,\eeq \beq \label{eq152} \hat\lambda(q,r) = \frac 1 3 w_0 r - \frac
f 2 \log \frac{\phi(q,r)}{\tilde\phi(q,r)} \qquad (when\;\rho > 0)\eeq

In terms of $\phi(q,r)$ and $\tilde\phi(q,r)$, the system \rif{148'},\rif{149'} becomes \barr
\label{eq153} -f\de{}{r}\phi(q,r) &= &\biggr(-\frac{f^2}{2\eta}\frac{\partial^2}{\partial q^2}+
V(q)-w_0+\nonumber \\ &&+\frac f r \log \frac{\phi(q,r)}{\tilde\phi(q,r)}\biggr)\phi(q,r) \earr
\beq \label{eq154} f\de{}{r}\tilde\phi(q,r) =
\biggr(-\frac{f^2}{2\eta}\frac{\partial^2}{\partial q^2}+ V(q)-w_0+\frac f r \log
\frac{\phi(q,r)}{\tilde\phi(q,r)}\biggr)\tilde\phi(q,r) \eeq We are interested to bounded
solutions of this system.

We note that, if we neglect the term $(f/r)\log \bigr(\phi(q,r)\bigr/\tilde\phi(q,r)\bigr)$,
the resulting equations \beq \label{eq155} -f\de{}{r}\phi_0(q,r)
=\biggr(-\frac{f^2}{2\eta}\frac{\partial^2}{\partial q^2}+ V(q)-w_0\biggr)\phi_0(q,r) \eeq \beq
\label{eq156} f\de{}{r}\widetilde{\phi_0}(q,r)
=\biggr(-\frac{f^2}{2\eta}\frac{\partial^2}{\partial q^2}+
V(q)-w_0\biggr)\widetilde{\phi_0}(q,r) \eeq can be easily solved. The eq.\rif{155}, which is a
diffusion equation, admits non negative solutions, bounded for every $r$, having the structure
\barr \label{eq157}
\phi_0(q,r)=c_0\psi_0(q)&+&c_1\psi_1(q)e^{-\frac{(w_1-w_0)}{f}r}+...\nonumber
\\ &+&c_n\psi_n(q)e^{-\frac{(w_n-w_0)}{f}r}+...\earr On the other hand,
\rif{156} admits only one solution, which is bounded for large $r$, that is \beq \label{eq158}
\widetilde{\phi_0}(q,r)= \widetilde{c_0}\psi_0(q) \eeq Now we take $c_0 = \widetilde{c_0} \;
(\neq 0)$, so that, for large $r$, \beq \label{eq159}\frac f r \log
\frac{\phi_0(q,r)}{\widetilde{\phi_0}(q,r)} = f \frac{c_1}{c_0} \frac 1 r
\frac{\psi_1(q)}{\psi_0(q)}e^{-\frac{(w_1-w_0)}{f}r}+O\biggr(\frac{e^{-\frac{(w_2-w_0)}{f}r}}{r}\biggr)\eeq
(assuming $c_1\neq0$). Due to \rif{159}, we can apply to the system \rif{153},\rif{154}, for
large $r$, the method of successive approximations, starting from $\phi_0(q,r)$ and
$\widetilde{\phi_0}(q,r)$. As a first improvement we obtain \barr \label{eq169} \phi(q,r)
&\simeq& \phi_0(q,r) + \int_r^{+\infty}\frac{\phi_0(q,r')}{r'}\log
\frac{\phi_0(q,r')}{\widetilde{\phi_0}(q,r')}dr' = \nonumber \\
&=& c_0 \psi_0(q)+c_1\psi_1(q)\biggr(1+\frac{f}{w_1-w_0}\cdot\frac
1 r-\nonumber
\\ &-&\frac{f^2}{(w_1-w_0)^2}\frac{1}{r^2}+...\biggr)e^{-\frac{(w_1-w_0)}{f}r}+...\earr
\barr \label{eq170} \widetilde{\phi}(q,r) \simeq c_0
\psi_0(q)&-&c_1\psi_1(q)\biggr(\frac{f}{w_1-w_0}\cdot\frac 1
r-\nonumber \\
&-&\frac{f^2}{(w_1-w_0)^2}\frac{1}{r^2}+...\biggr)e^{-\frac{(w_1-w_0)}{f}r}+...
\nonumber \\ \earr

The last dots represent terms of order $exp(-2(w_1-w_0)r/f)$ or $exp(-(w_2-w_0)/f)$, or smaller
than these. From \rif{151},\rif{169},\rif{170} we obtain that, for large $r$, \beq
\label{eq171} \rho(q,r) \simeq \psi_0^2(q)+c_1\psi_0(q)\psi_1(q)e^{-\frac{(w_1-w_0)}{f}r}+...
\,, \eeq where we have taken $c_0=+1$, for a proper normalization. Since $\psi_0(q)$ is the
state of the fundamental vacuum, \rif{171} describes a situation in which the physical system
of our field is static and confined within a region of the space, centered at the origin, of
radius $\sim f/(w_1-w_0)$. In such a situation we have that, besides the constant vacuum energy
density, there is a random energy density confined within a sphere of radius $\sim
f/(w_1-w_0)$.

We obtain analogous results if $c_1=0$ and $c_2$, for example, different from zero. We note
that, due to the invariance of our equations as regards spatial translations, the previous
considerations on static solutions can be applied to a generic sphere having center at an
arbitrary fixed point of the space. Then one can consider also several spheres, with centers
having distances larger than $2f/(w_1-w_0)$. The above results seem to have some appealing
aspects. Their interpretation and a further analysis of our equations for specific models will
be given elsewhere. We note that in the case of mechanical systems, where we have only the
parameter $t$, the balance between two competitive diffusion processes makes ineffective the
H-theorem. On the other hand, as we have seen, in the case of fields this balance does not
prevent that an H-theorem be operative in the space-like directions.

We add a brief comment on the dynamical behavior of our system \rif{125},\rif{128}. We can also
introduce for fields a quantity like the wave function of the mechanical systems. However, in
this case, the notion of wave function is associated to some covering of the space-time through
a family of space-like surfaces. Each such a covering determines a particular wave function. To
fix the ideas, let us consider the standard family of space-like surfaces, each having a
constant "time" $x_0$. To this family we can associate the wave function $\psi(q,x^0|\vec x)$,
parametrized by $\vec x$, given locally by \beq \psi(q,x^0|\vec x) = \rho^{1/2}(q,x^0|\vec x)
exp\bigr(\frac{\emph{i}}{f}\lambda^0(q,x^0|\vec x)\bigr)\nonumber \eeq

Due to sect.\ref{sec:level3}, the system \rif{125},\rif{128} can
be translated in the system \barr \label{eq172} && \emph{i}f
\de{}{x^0}\psi(q,x^0|\vec x)=-\frac{f^2}{2}\dpr
i\biggr(\eta^{ij}(q)\dpr j \psi(q,x^0|\vec x)\biggr) + \nonumber
\\ && +\biggr(\de{}{x^m}\lambda^m(q,x^0|\vec x)+\frac 1 2 \eta^{ij}(q)\dpr i \lambda_m(q,x^0|\vec x)\dpr j \lambda^m(q,x^0|\vec
x)\nonumber \\ && +V(q)\biggr)\psi(q,x^0|\vec x)=0\earr

\beq \label{eq173} \de{}{x^m}\bigr(\bar{\psi}\psi\bigr) + \dpr k
\bigr(\bar{\psi}\psi\eta^{kj}\dpr j \lambda_m\bigr) = 0 \qquad
(m=1,2,3)\eeq

We have then for $\psi(q,x^0|\vec x)$ a Schr\"{o}dinger equation, in which, besides the known
function $V(q)$, there is also a "self" potential, constrained by the equations \rif{173}. As a
result, even if we introduce the above wave function, the linearity found for mechanical system
is lost in the case of the fields.

However it is possible that linearity requires a mathematical object more general than
$\psi(q,x^0|\vec x)$.

We conclude here with some few remarks. The previous approach for scalar fields is based on a
particular well-defined probabilistic scheme, which makes use of ordinary functions. On the
other hand the standard approach of the quantum theory of fields, is based on a more general
probabilistic scheme which is of functional type, but ill-defined. However the probabilistic
distributions of our approach could be considered or imposed, as the appropriate marginal
distributions of functional distributions which are not well-defined.

Our approach requires some further developments in order to treat the case of singular
Lagrangians. An analysis of the Dirac fields is particularly important in order to see how the
previous probabilistic scheme for scalar fields must be modified. A further important problem
is to investigate the application of our approach to the case of the gravitational field.

\end{document}